\documentclass[useAMS,usenatbib]{mn2e}

\usepackage{natbib} %for references
\usepackage{multirow} %for multicolumns in tables
\usepackage[caption=false]{subfig} %for figures over multiple pages
\usepackage{enumerate} %for lists
\usepackage[T1]{fontenc} %for emphasis in titles
\usepackage{amsmath} %for equations
\usepackage{amssymb} %for equations
\usepackage{rotating} %for sideways figures
\usepackage{chngcntr} %for appendix figure numbers
\usepackage{appendix} %for appendices
\usepackage{book tabs} %for horizontal lines in tables
\usepackage[hyperindex, colorlinks, draft, citecolor=blue]{hyperref}  % for active links 

\def\reff@jnl#1{{\rm#1\/}}

\def\aj{\reff@jnl{AJ}}                  % Astronomical Journal
\def\araa{\reff@jnl{ARA\&A}}            % Annual Review of Astron and Astrophys
\def\apj{\reff@jnl{ApJ}}                % Astrophysical Journal
\def\apjl{\reff@jnl{ApJ}}               % Astrophysical Journal, Letters
\def\apjs{\reff@jnl{ApJS}}              % Astrophysical Journal, Supplement
\def\ao{\reff@jnl{Appl.Optics}}         % Applied Optics
\def\apss{\reff@jnl{Ap\&SS}}            % Astrophysics and Space Science
\def\aap{\reff@jnl{A\&A}}               % Astronomy and Astrophysics
\def\aapr{\reff@jnl{A\&A~Rev.}}         % Astronomy and Astrophysics Reviews
\def\aaps{\reff@jnl{A\&AS}}             % Astronomy and Astrophysics, Supplement
\def\azh{\reff@jnl{AZh}}                        % Astronomicheskii Zhurnal
\def\baas{\reff@jnl{BAAS}}              % Bulletin of the AAS
\def\jrasc{\reff@jnl{JRASC}}            % Journal of the RAS of Canada
\def\memras{\reff@jnl{MmRAS}}           % Memoirs of the RAS
\def\mnras{\reff@jnl{MNRAS}}            % Monthly Notices of the RAS
\def\pra{\reff@jnl{Phys.Rev.A}}         % Physical Review A: General Physics
\def\prb{\reff@jnl{Phys.Rev.B}}         % Physical Review B: Solid State
\def\prc{\reff@jnl{Phys.Rev.C}}         % Physical Review C
\def\prd{\reff@jnl{Phys.Rev.D}}         % Physical Review D
\def\prl{\reff@jnl{Phys.Rev.Lett}}      % Physical Review Letters
\def\pasp{\reff@jnl{PASP}}              % Publications of the ASP
\def\pasj{\reff@jnl{PASJ}}              % Publications of the ASJ
\def\qjras{\reff@jnl{QJRAS}}            % Quarterly Journal of the RAS
\def\skytel{\reff@jnl{S\&T}}            % Sky and Telescope
\def\solphys{\reff@jnl{Solar~Phys.}}    % Solar Physics
\def\sovast{\reff@jnl{Soviet~Ast.}}     % Soviet Astronomy
\def\ssr{\reff@jnl{Space~Sci.Rev.}}     % Space Science Reviews
\def\zap{\reff@jnl{ZAp}}                        % Zeitschrift fuer Astrophysik
\def\nat{\reff@jnl{Nature}}             % Nature 

\title[Planck Observations of M33]{Planck Observations of M33}

\author[Tibbs et al.]{C.T.~Tibbs$^{1}$\thanks{ESA Research Fellow}, F.P. Israel$^{2}$\thanks{E-mail: israel@strw.leidenuniv.nl}, R.J. Laureijs$^{1}$, J.A. Tauber$^{1}$, B. Partridge$^{3}$, M.W. Peel$^{4}$, \and L. Fauvet$^{5}$ \\
$^{1}$Scientific Support Office, Directorate of Science, European Space Research and Technology Centre (ESA/ESTEC), \\ Keplerlaan 1, 2201 AZ, Noordwijk, The Netherlands \\
$^{2}$Leiden Observatory, Leiden University, P.O. Box 9513, 2300 RA, Leiden, The Netherlands \\
$^{3}$Department of Astronomy, Haverford College, Haverford, PA 19041, USA \\
$^{4}$Departamento de F\'{i}sica Matematica, Instituto de F\'{i}sica, Universidade de S\~{a}o Paulo, S\~{a}o Paulo, Brazil \\
$^{5}$ARGANS Limited, Tamar Science Park, Plymouth, PL6 8BX, UK }

\begin{document}

\date{}

\pagerange{\pageref{firstpage}--\pageref{lastpage}} \pubyear{}

\maketitle

\label{firstpage}

%%%%%%% Abstract %%%%%%%%%%%%%%%%%%%%%%%%%%%%%%%%%%%%%%%%%%%%

\begin{abstract}
We have performed a comprehensive investigation of the global integrated flux density of M33 from radio to ultraviolet wavelengths, finding that the data between~$\sim$100\,GHz and 3\,THz are accurately described by a single modified blackbody curve with a dust temperature of $T_\mathrm{dust}$ = 21.67~$\pm$~0.30\,K and an effective dust emissivity index of $\beta_\mathrm{eff}$ = 1.35~$\pm$~0.10, with no indication of an excess of emission at millimeter/sub-millimeter wavelengths. However, sub-dividing M33 into three radial annuli, we found that the global emission curve is highly degenerate with the constituent curves representing the sub-regions of M33. We also found gradients in $T_\mathrm{dust}$ and $\beta_\mathrm{eff}$ across the disk of M33, with both quantities decreasing with increasing radius. Comparing the M33 dust emissivity with that of other Local Group members, we find that M33 resembles the Magellanic Clouds rather than the larger galaxies, i.e., the Milky Way and M31. In the Local Group sample, we find a clear correlation between global dust emissivity and metallicity, with dust emissivity increasing with metallicity. A major aspect of this analysis is the investigation into the impact of fluctuations in the Cosmic Microwave Background~(CMB) on the integrated flux density spectrum of M33. We found that failing to account for these CMB fluctuations would result in a significant over-estimate of $T_\mathrm{dust}$ by~$\sim$5\,K and an under-estimate of $\beta_\mathrm{eff}$ by~$\sim$0.4. 
\end{abstract}

%%%%%%% Key Words %%%%%%%%%%%%%%%%%%%%%%%%%%%%%%%%%%%%%%%%%%%%


%%%%%%% Introduction %%%%%%%%%%%%%%%%%%%%%%%%%%%%%%%%%%%%%%%%%

\section{Introduction}
\label{Sec:Intro}

In the region between high-frequency radio waves ($\nu$\,$\gtrsim$\,10\,GHz) and long-wavelength infrared~(IR) emission ($\lambda$\,$\gtrsim$\,100\,$\mu$m), thermal radiation from interstellar dust and ionized gas, as well as non-thermal synchrotron radiation, all contribute to the emission from cosmic objects. By unravelling the various contributions, we may obtain information on the ionising stars and the properties of interstellar dust in a variety of galactic environments. The observations provided by the \textit{Planck} mission~\citep{Tauber:10} allow us to sample the poorly observed far-IR to millimetre~(mm) gap in the continuum emission spectrum of objects such as entire galaxies. 

In the past, attempts have been made to extrapolate the incomplete IR continuum flux density spectrum~(frequently, but incorrectly, referred to as a spectral energy distribution or SED\footnote{An SED is a plot of energy as a function of frequency or wavelength, i.e., $\nu S_{\nu}$ vs $\nu$, or $\lambda S_\lambda$ vs $\lambda$, while a flux density spectrum is a plot of flux density as a function of frequency or wavelength i.e., $S_{\nu}$ vs $\nu$, or $S_{\lambda}$ vs $\lambda$.}) cutting off somewhere between 100 and 160\,$\mu$m (\textit{IRAS, Spitzer Space Telescope}) by assuming a single effective integrated dust emissivity index of $\beta_\mathrm{eff}$ = 2 for the Rayleigh-Jeans extrapolation. Often, the values actually measured at wavelengths around 1\,mm significantly exceed such extrapolated flux densities. This so-called ``millimetre excess'' was readily interpreted as evidence for a large mass of colder dust~\citep[see, for instance,][]{Galliano:05}. However, both the increased far-IR wavelength coverage~(up to 500\,$\mu$m) of the \textit{Herschel Space Observatory} and the results of terrestrial laboratory experiments have subsequently indicated that the actual value of $\beta_\mathrm{eff}$ is generally $<$\,2. As a consequence, both the historic millimetre excess and the implied large mass of colder dust, can be reduced to an artefact of interpretation, and effectively disappear.

Thus far, reliable and complete continuum flux density spectra ranging from the radio to the mid-IR or even optical wavelengths, well-sampling the mm to far-IR range, have been published for a variety of Milky Way sources, but only for a few galaxies beyond. The wavelength coverage of \textit{Planck} renders extrapolation superfluous; the value of $\beta_\mathrm{eff}$ can be measured directly. This measurement is complicated by the degeneracy between $\beta_\mathrm{eff}$ and the dust temperature, $T_\mathrm{dust}$, derived from flux densities that have finite instrumental noise: the two parameters are inversely correlated~\citep[e.g.,][]{Shetty:09, Juvela:12a, Juvela:12b}. Nevertheless, the \textit{Planck} spectra of the Local Group galaxies, the Large and Small Magellanic Clouds~\citep[LMC and SMC;][]{Planck_Early_Results_XVII:11}, and M31~\citep{Planck_Intermediate_Results_XXV:15}, imply galaxy-wide effective dust emissivities well below two. Similar emissivities have been found for other nearby galaxies~\citep{Planck_Early_Results_XVI:11}.

Surprisingly, the complete flux density spectra of the LMC and SMC, incorporating \textit{WMAP} and \textit{COBE} data, published by~\citet{Israel:10} and interpreted by~\citet{Bot:10}, do show a pronounced excess of emission at mm to cm wavelengths. This ``new'' excess emission is not to be confused with the apparent historical millimetre excess discussed earlier as it does not result from an arbitrary assumption on the dust emissivity, but is a well-sampled spectral feature. Its existence was confirmed by~\citet{Planck_Early_Results_XVII:11}, who explained the observed excess in the LMC as a fluctuation of the Cosmic Microwave Background~(CMB), but admitted to the presence of a significant intrinsic excess in the SMC. \citet{Draine:12} proposed that this intrinsic excess in the SMC could be explained if the interstellar dust includes magnetic nanoparticles, emitting magnetic dipole radiation resulting from the thermal fluctuations in the magnetisation.

Perhaps relatedly, an excess of emission at longer (cm) wavelengths has also been observed in many environments within the Milky Way~\citep[see][and references within]{Planck_Intermediate_Results_XV:14}. This cm excess, more commonly known as anomalous microwave emission (AME), is typically observed at frequencies around 30\,GHz~(or wavelengths of 1\,cm), is observed to be highly correlated with the IR dust emission~\citep[e.g.,][]{Casassus:06, Tibbs:10, Tibbs:13, Planck_Intermediate_Results_XV:14}, and is believed to be due to electric dipole radiation from very small rapidly spinning dust grains~\citep{Draine:98}.

\begin{figure*}
\begin{center}
\includegraphics[angle=0,scale=0.4555]{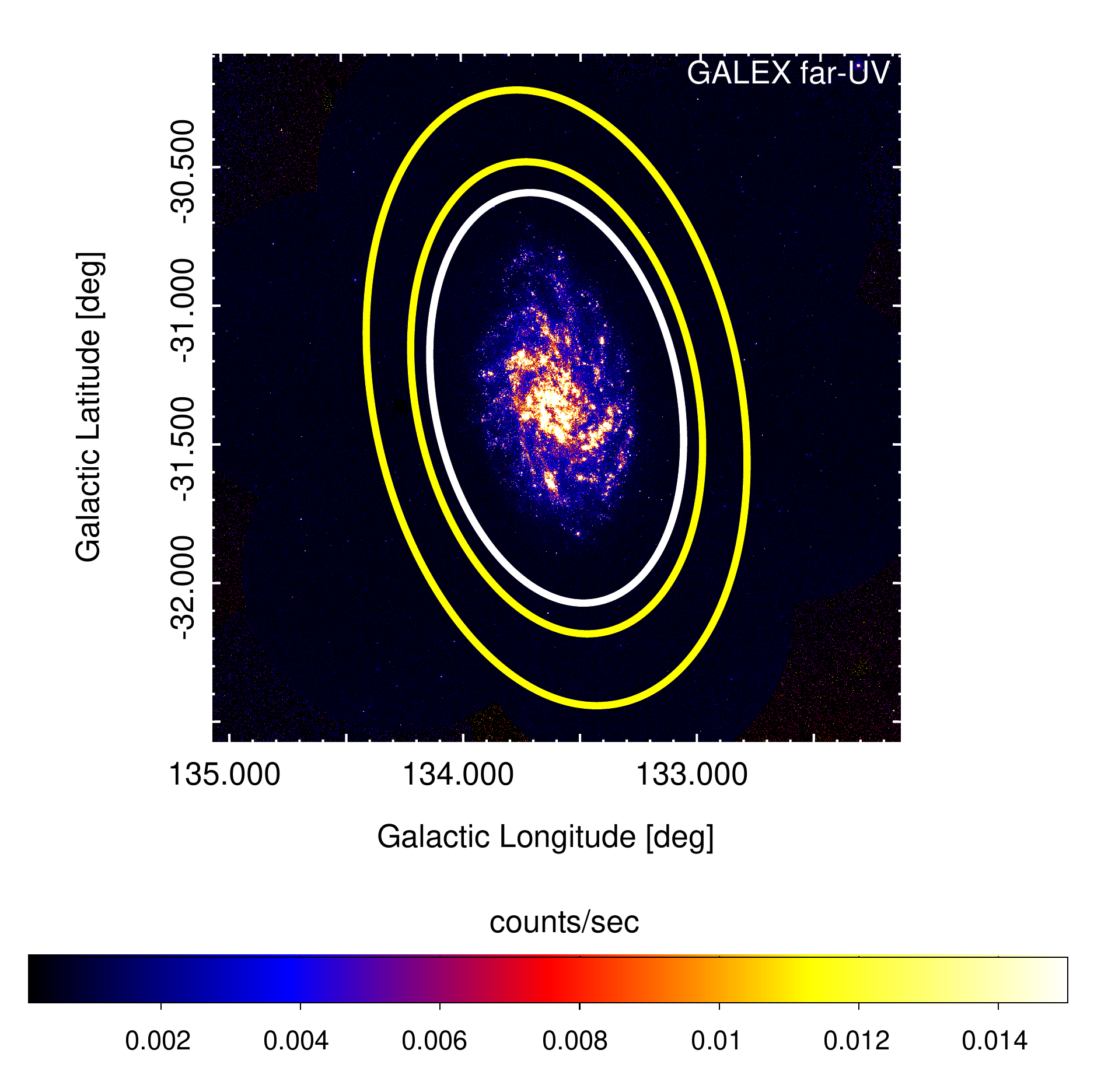} 
\includegraphics[angle=0,scale=0.4555]{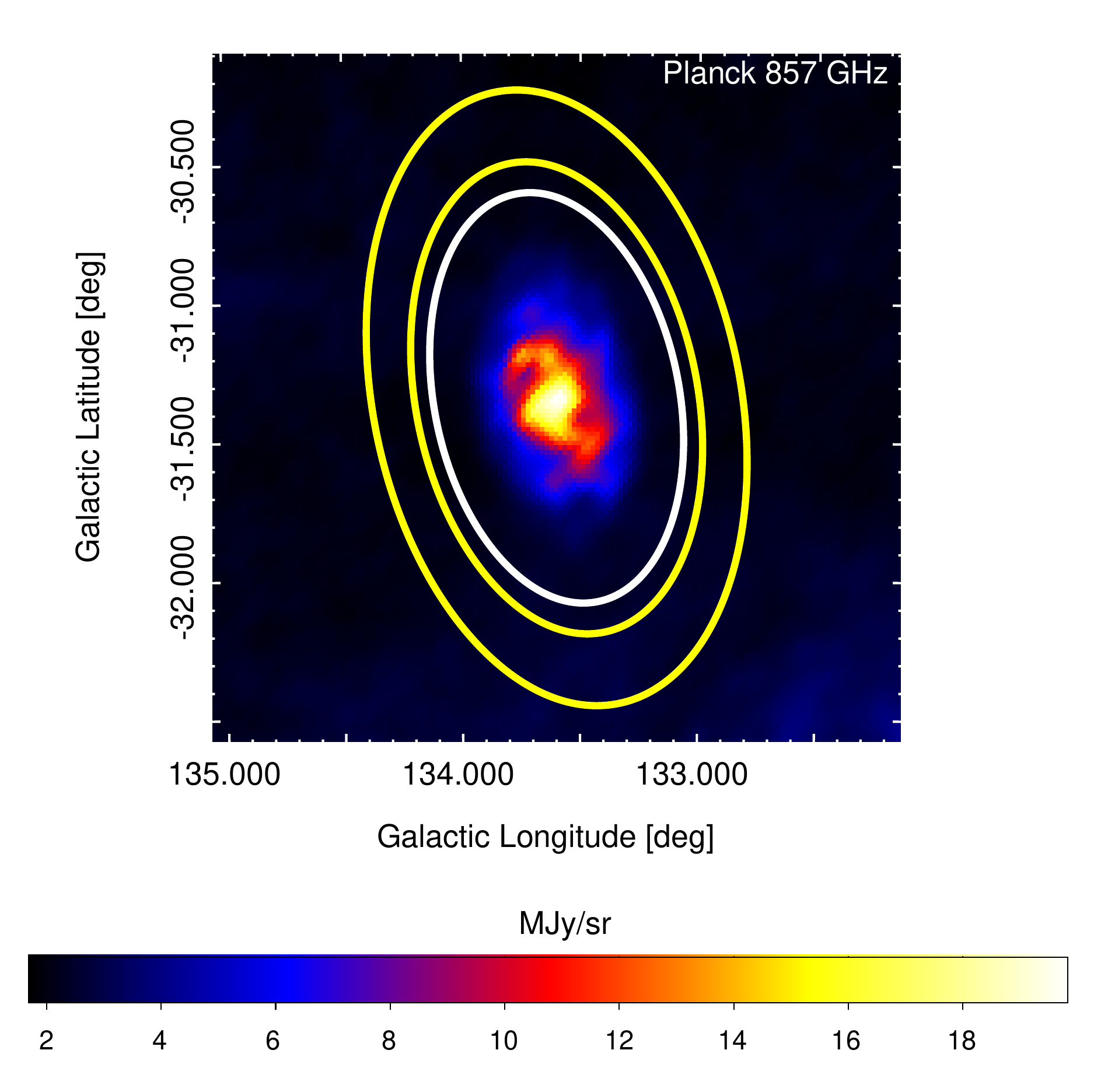} 
\end{center}
\vspace{-0.3cm}
\caption{GALEX far-UV~(\textit{left}) and \textit{Planck} 857\,GHz~(\textit{right}) map of M33. The aperture (white ellipse) and background annulus (green ellipses) used to compute the integrated emission in M33 are shown.}
\label{Fig:M33_PLCK857}
\end{figure*}

In this paper we present a study of the small Local Group spiral galaxy M33, using the most recent \textit{Planck} data along with data from the literature, to produce a comprehensive continuum flux density spectrum from radio to ultraviolet~(UV) wavelengths. We profit from the fact that, due to its proximity~\citep[$d$ = 840~kpc,][]{Freedman:91} and modest dimensions (approximately 70\,arcmin~$\times$~40\,arcmin - see Fig.~\ref{Fig:M33_PLCK857}), M33 is an exceedingly well-studied object. In this analysis we will specifically address: (a) the shape of the Rayleigh-Jeans spectrum, (b) the magnitude of the effective dust emissivity spectral index, $\beta_\mathrm{eff}$, and (c) possible differences between the inner and outer regions of M33. Since the flux density spectra of both individual interstellar clouds and entire galaxies have a minimum close to the peak of the CMB, at these frequencies the CMB emission typically exceeds the interstellar contribution. Thus, our results depend critically on the reliability of the CMB subtraction. For this reason, we will pay special attention to an analysis of the CMB fluctuations, as these dominate the M33 spectrum at mm wavelengths. 

This paper is organised as follows. In Section~\ref{Sec:Data} we describe the data used in this analysis, while in Section~\ref{Sec:Analysis} we produce a global continuum flux density spectrum for M33, accounting for contributions from both CMB fluctuations and CO line emission. We also spatially decompose M33 into three annuli, producing a flux density spectrum for each. In Section~\ref{Sec:Discussion} we discuss the results of our work, and we present our conclusions in Section~\ref{Sec:Conclusions}.

%%%%%%% Observations %%%%%%%%%%%%%%%%%%%%%%%%%%%%%%%%%%%%%%%%%

\section{Data}
\label{Sec:Data}

\begin{table}
\begin{center}
\caption{Characteristics of the far-IR/sub-mm data used in this analysis including the reference frequency, $\nu_\mathrm{ref}$, the angular resolution, $\theta$, and the photometric uncertainty, $\epsilon_\mathrm{phot}$.}
\begin{tabular}{lcccc}
\hline
Telescope/Instrument & $\nu_\mathrm{ref}$ & $\theta$ & $\epsilon_\mathrm{phot}$ \\
 & (GHz) & (arcmin) & \\
\hline
\hline

\textbf{\textit{Planck}} 	&			&		& \\
\hspace{1cm}LFI030		& 28.4		& 32.3	& 1$\%$ \\	
\hspace{1cm}LFI044		& 44.1		& 27.1	& 1$\%$ \\	
\hspace{1cm}LFI070		& 70.4		& 13.3	& 1$\%$ \\	
\hspace{1cm}HFI100		& 100		& 9.7		& 1$\%$ \\	
\hspace{1cm}HFI143		& 143		& 7.3		& 1$\%$ \\
\hspace{1cm}HFI217		& 217		& 5.0		& 1$\%$ \\	
\hspace{1cm}HFI353		& 353		& 4.9		& 1$\%$ \\	
\hspace{1cm}HFI545		& 545		& 4.8		& 7$\%$ \\	
\hspace{1cm}HFI857		& 857		& 4.6		& 7$\%$ \\	
\\
\textbf{\textit{Herschel}}			&		    	&		& \\
\hspace{1cm}SPIRE 500$\mu$m	&	600	    	& 0.60	& 10$\%$ \\
\hspace{1cm}SPIRE 350$\mu$m	&	857     	& 0.41	& 10$\%$ \\
\hspace{1cm}SPIRE 250$\mu$m	&	1200  	& 0.30	& 10$\%$ \\
\hspace{1cm}PACS 160$\mu$m	&	1870  	& 0.19	& 15$\%$ \\
\hspace{1cm}PACS 100$\mu$m	&	3000  	& 0.12	& 15$\%$ \\
\hspace{1cm}PACS 70$\mu$m		&	4280  	& 0.09	& 15$\%$ \\
\\
\textbf{\textit{IRAS}/IRIS}	&				&		& \\
\hspace{1cm}100$\mu$m	&	3000			& 4.3		& 13.5$\%$ \\
\hspace{1cm}60$\mu$m	&	5000       		& 4.0	 	& 10.4$\%$ \\
\hspace{1cm}25$\mu$m	&	12000 	 	& 3.8		& 15.1$\%$ \\
\hspace{1cm}12$\mu$m	&	25000  		& 3.8		& 5.1$\%$ \\
\\
\textbf{\textit{Spitzer}}		&			&		& \\
\hspace{1cm}MIPS 24$\mu$m	& 12500  		& 0.10	& 10$\%$ \\
\hspace{1cm}IRAC 8$\mu$m	& 37500  		& 0.03	& 10$\%$ \\
\hspace{1cm}IRAC 5.8$\mu$m	& 51700  		& 0.03	& 10$\%$ \\
\hspace{1cm}IRAC 4.5$\mu$m	& 66600  		& 0.03	& 10$\%$ \\
\hspace{1cm}IRAC 3.6$\mu$m	& 83300  		& 0.03	& 10$\%$ \\

\hline
\label{Table:Data}
\end{tabular}
\end{center}
\vspace{-0.6cm}
%Table notes here
\end{table}

\subsection{Planck}
\label{Subsec:Planck}

The \textit{Planck} mission~\citep{Tauber:10, Planck_Early_Results_I:11} was the third cosmological satellite mission to observe the entire sky in a series of wide spectral passbands~($\Delta \nu/\nu\sim$0.3--0.6) designed to sample the CMB. It measured the emission from the sky with the Low Frequency Instrument (LFI) at 28.4, 44.1, and 70.4\,GHz (1.0--0.4\,cm) with amplifiers cooled to 20\,K between 2009 and 2013, and the High Frequency Instrument (HFI) at 100, 143, 217, 353, 545, and 857\,GHz (3.0--0.35\,mm), with bolometers cooled to 0.1\,K between 2009 and early 2012 (cf., Table~\ref{Table:Data}). 

In this paper, we use the most recent \textit{Planck} 2015 ``full'' data release~\citep{Planck_2015_Results_I:16}. These data cover the full mission from 12 August 2009 to 23 October 2013 and are available to download from the \textit{Planck} Legacy Archive.\footnote{http://pla.esac.esa.int/pla/} The full LFI/HFI data processing and calibration procedures are described in \citet{Planck_2015_Results_II:16, Planck_2015_Results_III:16, Planck_2015_Results_IV:16, Planck_2015_Results_V:16, Planck_2015_Results_VI:16, Planck_2015_Results_VII:16, Planck_2015_Results_VIII:16} with an overview provided in~\citet{Planck_2015_Results_I:16}. The \textit{Planck} full-sky maps are provided in \textsc{HEALPix} format~\citep{Gorski:05}, but for this analysis we extracted 2D projected maps centred on M33 using the \textsc{Gnomdrizz} package~\citep{Paradis:12}, which accurately conserves the photometry during the data re-pixelization. After performing this extraction for each of the nine \textit{Planck} frequency maps, the maps were converted from units of CMB temperature to MJy\,sr$^{-1}$ using the coefficients described in~\citet{Planck_2013_Results_IX:14}. The nine \textit{Planck} frequency maps are displayed in the left column of Fig.~\ref{Fig:Planck_maps_all}.

The 2015 \textit{Planck} data have been calibrated on the orbital modulation of the ``cosmological dipole'', resulting in extremely high~(sub-percent) calibration accuracy~\citep[see table 1 from][]{Planck_2015_Results_I:16}. However, it is important to note that the quoted accuracies are appropriate for diffuse emission at large angular scales, where the calibration signal appears. Additional uncertainties apply at smaller angular scales. For relatively compact sources such as M33, the main additional contributors are related to colour correction and to beam uncertainty. Colour correction uncertainties depend on the spectral shape of the source~\citep{Planck_2015_Results_II:16, Planck_2015_Results_VII:16}, while the beam uncertainties depend on angular scale~\citep{Planck_2015_Results_IV:16, Planck_2015_Results_VII:16}; for M33 we conservatively assume the entire solid angle uncertainty to be applicable. Combining these uncertainties in quadrature, we conservatively assume a photometric uncertainty of 7\,\% for the 545 and 857\,GHz bands, and 1\,\% for the other seven bands. Ultimately, the uncertainty on the flux determination of compact sources is limited by fluctuations in both the physical backgrounds and foregrounds rather than photometry errors.

\subsection{Herschel}
\label{Subsec:Herschel}

M33 was mapped with \textit{Herschel} within the framework of the open-time key programme as part of the HerM33es KPOT\_ckrame01\_1~\citep{Kramer:10} and OT2\_mboquien\_4~\citep{Boquien:15} proposals. The~\citet{Kramer:10} observations (observation ID 1342189079 and 1342189080) were performed simultaneously with the PACS~(100 and 160\,$\mu$m) and SPIRE~(250, 350, and 500\,$\mu$m) instruments in parallel mode in two orthogonal directions to map a region of approximately 90\,arcmin~$\times$~90\,arcmin. The~\citet{Boquien:15} observations (observation ID 1342247408 and 1342247409) were performed solely with the PACS instrument in two orthogonal directions at 70 and 160\,$\mu$m, and covered a smaller area of approximately 50\,arcmin~$\times$~50\,arcmin. These maps have spatial resolutions ranging from approximately 6 to 11\,arcsec for the PACS maps and approximately 18 to 37\,arcsec for the SPIRE maps. 

The fully reduced maps were made publicly available by the HerM33es team as \textit{Herschel} User Provided Data Products, and we downloaded these data from the \textit{Herschel} Science Archive.\footnote{http://archives.esac.esa.int/hsa/whsa/} Full details of the PACS and SPIRE data reduction and map-making are described in detail by~\citet{Boquien:11, Boquien:15} and~\citet{Xilouris:12}, respectively. Following~\citet{Boquien:11} and~\citet{Xilouris:12}, we assume a 15 and 10\,\% photometric uncertainty on the extended emission in these PACS and SPIRE maps, respectively.

\subsection{IRAS}
\label{Subsec:IRAS}

The original \textit{IRAS} measurements of M33 were presented and discussed by~\citet{Rice:90}. In this analysis we use the Improved Reprocessing of the \textit{IRAS} Survey~\citep[IRIS;][]{Miville-Deschenes:05} data for all four \textit{IRAS} bands at 12, 25, 60, and 100\,$\mu$m. These data have been reprocessed resulting in an improvement in the zodiacal light subtraction, destriping, and calibration. We use the photometric uncertainties estimated by~\citet{Miville-Deschenes:05} of 5.1, 15.1, 10.4, and 13.5\,\% for the 12, 25, 60, and 100\,$\mu$m bands, respectively.

\subsection{Spitzer}
\label{Subsec:Spitzer}

\textit{Spitzer} mapped M33 as part of the Guaranteed Time Observations (PID 5, PI. R. Gehrz) and we use the MIPS 24\,$\mu$m data along with the IRAC 8.0, 5.8, 4.5, and 3.6\,$\mu$m data. The IRAC observations have spatial resolutions of~$\sim$2.0\,arcsec, while the MIPS 24\,$\mu$m map has a resolution of 6\,arcsec. A detailed description of the \textit{Spitzer} photometry of M33 is provided by~\citet{Verley:07, Verley:09}. For this analysis, we downloaded the data from the \textit{Spitzer} Heritage Archive\footnote{http://sha.ipac.caltech.edu/applications/Spitzer/SHA/} and reprocessed the data by subtracting the contribution from the zodiacal light, applying the extended emission correction, mosaicking the data, performing an overlap correction, and subtracting the brightest point sources. This processing was performed using \textsc{mopex} and \textsc{apex} in a similar manner to that discussed in~\citet{Tibbs:11}, and we assume a calibration uncertainty of 10$\%$ on these maps.

%%%%%%% Analysis And Results %%%%%%%%%%%%%%%%%%%%%%%%%%%%%%%%%%%%%%%%%

\section{Analysis And Results}
\label{Sec:Analysis}

\begin{figure}
\begin{center}
\includegraphics[angle=0,scale=0.65]{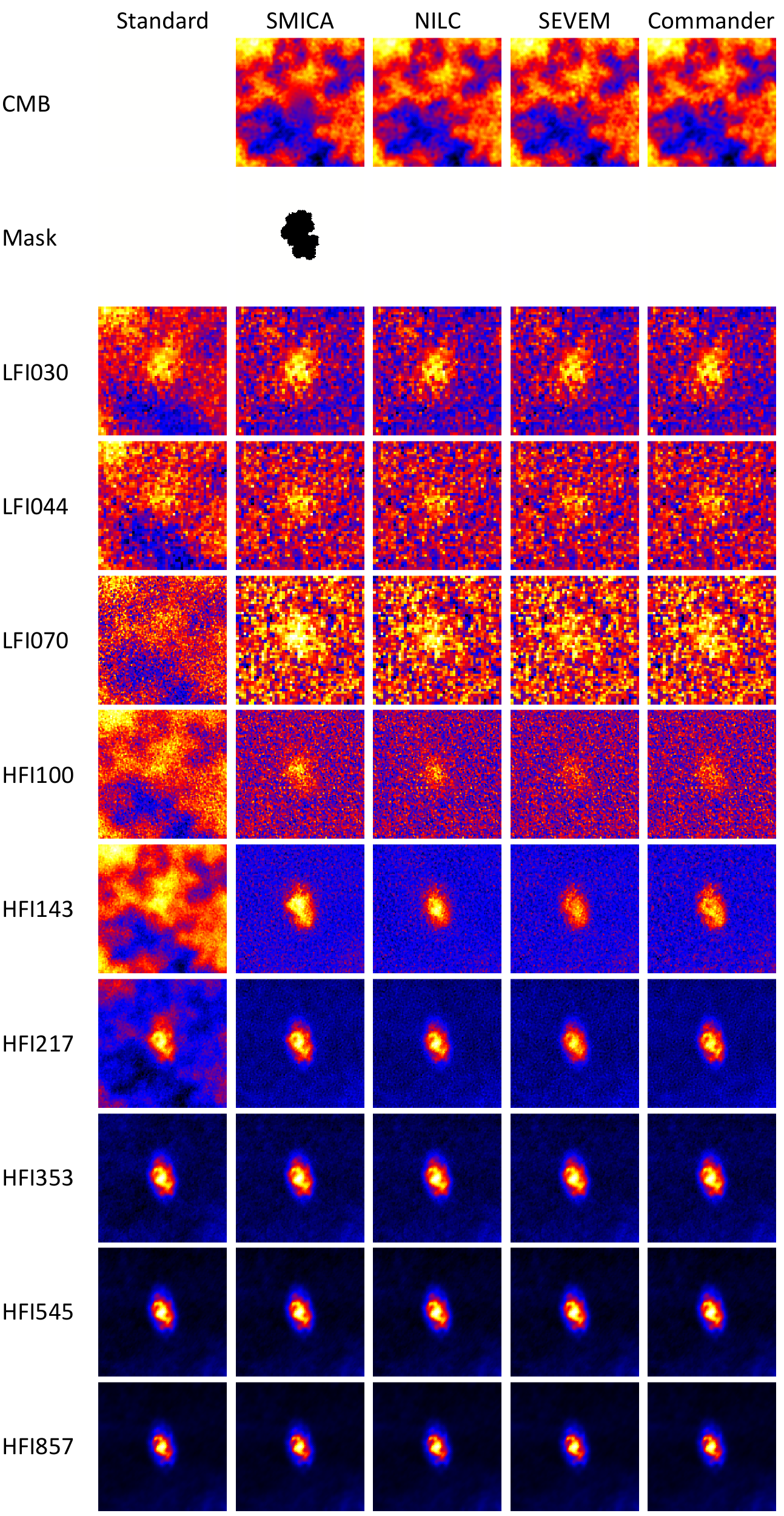}
\end{center}
\vspace{-0.3cm}
\caption{All of the \textit{Planck} maps used in this analysis. The first column shows the nine standard \textit{Planck} frequency maps, which contain a contribution from the CMB, while the second, third, fourth, and fifth columns show the CMB map, the CMB mask, and the nine corresponding frequency maps which have had the CMB contribution subtracted using the \texttt{SMICA}, \texttt{NILC}, \texttt{SEVEM}, and \texttt{Commander} methods, respectively. Each individual map is 2.5$^\circ$ across and orientated in the Galactic coordinate system as defined in Fig.~\ref{Fig:M33_PLCK857}.}
\label{Fig:Planck_maps_all}
\end{figure}

\subsection{Aperture photometry}
\label{Sec:Photometry}

In order to determine the integrated flux densities for M33 we applied aperture photometry to the datasets summarised in Section~\ref{Sec:Data}. For our aperture photometry analysis we used an elliptical aperture with a semi-major axis of 45\,arcmin~(11\,kpc at the distance of M33), a semi-major to semi-minor axis ratio of $10^{0.23}$~\citep{Paturel:03}, and a position angle of 22.5\,degrees with respect to an equatorial reference frame~\citep{Kramer:10}, centred on M33 ($\ell$~=~133.60$^{\circ}$, $b$~=~$-$31.34$^{\circ}$) as indicated in Fig.~\ref{Fig:M33_PLCK857}. The size of the aperture was selected based on computing the integrated flux density in apertures of increasing size to determine when the computed flux density stopped growing. The unrelated background and foreground emission was estimated within an elliptical annulus with inner and outer semi-major axes of 1.15 and 1.50 times the aperture semi-major axis, respectively.

The estimated uncertainty on the computed flux densities is a combination of the photometric uncertainty, $\epsilon_\mathrm{phot}$, for which we have adopted the values listed in Table~\ref{Table:Data}, and the flux measurement uncertainty, $\epsilon_\mathrm{bg}$, which contains two terms: the first term is the variance in the aperture flux, and the second term is due to the background/foreground subtraction. Both terms are estimated from the variance in the annulus surrounding the source, computed as

\begin{equation}
\epsilon_\mathrm{bg} = \sigma_\mathrm{bg} \left[ N_\mathrm{aper} + \frac{\pi N_\mathrm{aper}^{2}}{2 N_\mathrm{bg}} \right]^{0.5} ,
\label{equ:unc1}
\end{equation}

\noindent
\citep[see also][]{Laher:12, Hermelo:16}, where $\sigma_\mathrm{bg}$ is the standard deviation of the pixels within the annulus, and $N_\mathrm{aper}$ and $N_\mathrm{bg}$ are the number of pixels within the aperture and the annulus, respectively. The total uncertainty on the computed flux densities is then estimated to be

\begin{equation}
\epsilon = \sqrt{\epsilon_\mathrm{phot}^{2} + \epsilon_\mathrm{bg}^{2}} .
\label{equ:unc3}
\end{equation}

The aperture photometry described above takes into account the \textit{average} CMB contribution, but not the effect of the CMB fluctuations, which may have a considerable impact on estimates of the M33 integrated flux density. Therefore, before determining the intrinsic M33 flux density spectrum, we need to consider the effect of the CMB on the shape of the spectrum.

%%%%%%% CMB Contribution %%%%%%%%%%%%%%%%%%%%%%%%%%%%%%%%%%%%%%%%%

\subsection{CMB contribution}
\label{Subsec:CMB_Contribution}

\begin{figure*}
\begin{center}
\includegraphics[angle=0,scale=1.0]{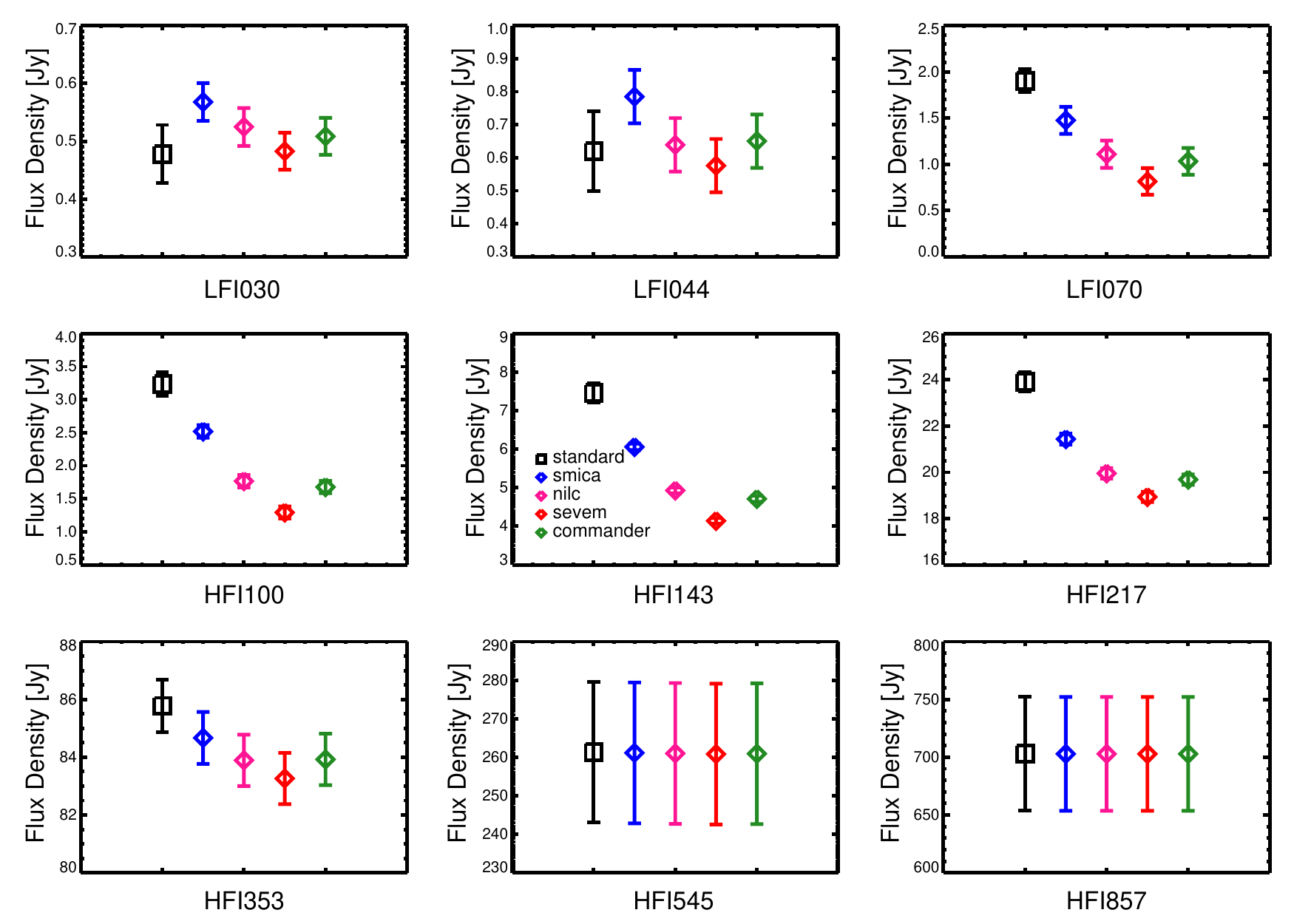}
\end{center}
\vspace{-0.3cm}
\caption{Flux densities in the nine \textit{Planck} channels for the standard maps, along with the CMB-subtracted maps.}
\label{Fig:CMB_Flux_Dens}
\end{figure*}

At the galactic latitude of M33, Milky Way foreground emission is relatively weak, even at the higher frequencies, and the main background affecting the source flux measurements is the CMB itself, especially at the lower observed frequencies. As can be seen from the left column in Fig.~\ref{Fig:Planck_maps_all}, M33 only starts to become clearly visible above the CMB at frequencies $\gtrsim$\,143\,GHz. \textit{Planck} has extracted the CMB over the whole sky using four different methods: \texttt{SMICA}, a non-parametric method that computes the CMB map by linearly combining all of the \textit{Planck} maps with weights that vary with multipole in the spherical harmonic domain; Needlet Internal Linear Combination~(\texttt{NILC}), which produces a CMB map using the \textit{Planck} maps between 44 and 857\,GHz by applying the Internal Linear Combination technique in the needlet (wavelet) domain; \texttt{SEVEM}, which estimates a CMB map based on linear template fitting in the map domain using internal templates constructed using the \textit{Planck} data; and \texttt{Commander}, which is a Bayesian parametric method that models all of the astrophysical signals in the map domain~\citep{Planck_2015_Results_IX:16}. 

It is important to note that in these CMB estimations, bright sources in the input maps are masked and the resulting CMB maps contain ``inpainted'' values within the masked areas. These inpainted values are good estimations of the CMB near the edge of the masked area (to preserve continuity), but are only statistically representative of the CMB within the mask. For this reason, the CMB masks employed by each of the four CMB separation techniques, along with the resulting CMB maps for the vicinity of M33, are displayed in Fig.~\ref{Fig:Planck_maps_all}, where it is apparent that M33 was only masked for the \texttt{SMICA} analysis. In addition to providing maps of the CMB, the \textit{Planck} Legacy Archive also provide maps containing only foreground emission~(i.e., the frequency maps with the CMB contribution subtracted). Since there are four different CMB maps, this produces four sets of CMB-subtracted maps. As recommended by~\citet{Planck_2015_Results_IX:16}, it is not advisable to produce an analysis that depends solely on a single component separation algorithm, and therefore we investigate the impact of the CMB by incorporating all four of the CMB-subtracted datasets in our analysis. As for the standard \textit{Planck} maps discussed in Section~\ref{Subsec:Planck}, we used \textsc{Gnomdrizz} to extract 2D projected maps of the CMB-subtracted maps and subsequently converted them into MJy\,sr$^{-1}$. These CMB-subtracted maps are displayed in Fig.~\ref{Fig:Planck_maps_all}, where it is clear to see variations between the maps, reflecting the different approaches used to estimate the CMB. The differences are most obvious at the lower frequencies $\lesssim$\,200\,GHz, where the CMB and its fluctuations are strongest. 

The \textit{Planck} consortium recommends using the \texttt{SMICA} CMB map at small angular scales, as it results in the lowest foreground residuals. However, in this case, since M33 has been masked, it is the least reliable of the four CMB-subtraction techniques. Nonetheless, since all four of the CMB-subtracted maps are in principle statistically indistinguishable, we keep all four of them in our analysis.

In order to quantify the effect of the CMB as a contaminant of the integrated emission within M33, we compared the flux densities~(estimated using aperture photometry as described in Section~\ref{Sec:Photometry}) in each of the nine \textit{Planck} bands for the five different datasets~(the standard \textit{Planck} frequency maps and the \texttt{SMICA}, \texttt{NILC}, \texttt{SEVEM}, and \texttt{Commander} CMB-subtracted maps), which we plot in Fig.~\ref{Fig:CMB_Flux_Dens}. These plots clearly illustrate how the CMB impacts the estimate of the integrated flux density. Not only are there differences between the flux density computed using the standard (non CMB-subtracted) maps, but there are also variations between the different CMB estimation methods. These variations are systematic as they only depend on the map frequency and input CMB amplitude, with \texttt{SMICA} and \texttt{SEVEM} producing the highest and lowest flux densities, respectively, and they also reflect that M33 is masked for the \texttt{SMICA} analysis, but not for the other three methods. Fig.~\ref{Fig:CMB_Flux_Dens} also quantitatively shows what can be seen in Fig.~\ref{Fig:Planck_maps_all}, which is that the CMB contamination is non-negligible at $\nu$\,$\lesssim$\,217\,GHz, while at higher frequencies the emission from M33 is dominant and hence the CMB contribution becomes increasing negligible.

%%%%%%% CO Contribution %%%%%%%%%%%%%%%%%%%%%%%%%%%%%%%%%%%%%%%%%

\subsection{Contamination by CO line emission}
\label{Subsec:CO_Contribution}

The wide passbands of the \textit{Planck} instruments cover the rest frequencies of various molecular lines. Emission from these lines will contaminate the measured broadband continuum flux densities, causing the latter to be overestimated. As discussed by~\citet{Planck_2013_Results_XIII:14}, Galactic line emission from carbon monoxide~(CO) is strongly detected in the \textit{Planck} bands. Specifically, only the $J$=1-0, $J$=2-1, and $J$=3-2 transitions of $^{12}$CO need to be considered, as only they are strong enough to have a significant effect on the \textit{Planck} HFI100, HFI217, and HFI353 bands, respectively. Although the \textit{Planck} data themselves have been used to produce maps of the $^{12}$CO emission~\citep{Planck_2015_Results_X:16}, these maps were produced for the velocity range of the Milky Way, which is substantially different to that of M33. We have inspected these maps and determined that they are largely unsuitable for this analysis. However, we can still investigate the magnitude of possible CO contamination by using the $^{12}$CO maps obtained with ground-based telescopes. M33 has been observed by \citet{Heyer:04}, who used the Five College Radio Astronomy Observatory~(FCRAO) 14\,m telescope to map the $J $=1-0 CO emission, while \citet{Druard:14} used the $J$=2-1 CO maps observed using the Institut de Radioastronomie Millim\'{e}trique~(IRAM) 30\,m telescope. Based on these data, we computed a flux density of the $J$=1-0 CO emission in the HFI100 band to be $\sim$0.1\,Jy, which when compared to the total flux densities estimated in the Planck 100\,GHz band accounts for~$\lesssim$\,9\,\% of the emission. Likewise, from the IRAM data~\citep[see also][]{Hermelo:16} we estimated a $J$=2-1 CO flux density in the HFI217 band of~$\sim$0.7\,Jy, which accounts for~$\lesssim$\,4\,\% of the emission in that band. Finally, the compilation of CO line ratios in spiral galaxies~(Israel et al., in prep.) suggests that the integrated brightness temperature ratio between the $J$=3-2 and $J$=1-0 lines is~$\sim$0.7, predicting a corresponding $J$=3-2 CO flux density of~$\sim$1.9\,Jy, which is $\lesssim$\,3\,\% of the emission in the HFI353 band.

Therefore, we use these flux densities to correct for the contribution from CO line emission. Throughout the rest of this analysis, we include small flux density corrections by subtracting the estimated CO flux density from the observed flux densities at 100, 217, and 353\,GHz in order to determine the intrinsic M33 continuum flux density spectrum as accurately as possible.

%%%%%%% SED %%%%%%%%%%%%%%%%%%%%%%%%%%%%%%%%%%%%%%%%%%%%%%%%

\subsection{Global continuum flux density spectrum and spectral energy distribution}
\label{Subsec:Full_SED}

To produce the global continuum flux density spectrum for M33, we performed aperture photometry on the \textit{Planck}, \textit{Herschel}, IRIS, and \textit{Spitzer} maps at full angular resolution. The resulting flux densities are listed in Table~\ref{Table:M33_Flux_Dens}. As discussed in Section~\ref{Subsec:CMB_Contribution}, to account for the effect of the CMB, we computed the flux densities for the standard \textit{Planck} maps along with the CMB-subtracted \textit{Planck} maps. The flux densities listed in Table~\ref{Table:M33_Flux_Dens} have also been corrected for CO contamination as described in Section~\ref{Subsec:CO_Contribution}.

For a proper analysis of the M33 continuum spectrum, we expand its frequency~(wavelength) range by adding results available in the literature, in addition to those listed in Table~\ref{Table:M33_Flux_Dens}. These include the radio flux densities from the compilation by~\citet{Israel:92} and from~\citet{Tabatabaei:07a}, as well as the 2MASS $J$, $H$, and $K_{S}$ band flux densities by~\citet{Jarrett:03}, the $U$, $B$, $V$ values from~\citet{DeVaucouleurs:91}, and the \textit{GALEX} far- and near-UV flux densities by~\citet{Lee:11}. The resulting flux density spectra, with and without CMB-subtraction, are displayed in Fig.~\ref{Fig:M33_Full_CFD}, while the corresponding SEDs, obtained by multiplying each flux density by its frequency, are shown in Fig.~\ref{Fig:M33_Full_SED}.

\begin{table*}
\begin{center}
\caption{Integrated flux densities of M33 both with and without subtracting the contribution from the CMB used to produce the continuum flux density spectra displayed in Fig.~\ref{Fig:M33_Full_CFD}. All of the flux densities estimated in this analysis have been colour-corrected and the HFI100, HFI217, and HFI353 channels have been corrected for CO contamination.}
\begin{tabular}{lcccccccc}
\hline
Instrument & Frequency & Wavelength & Flux Density & \multicolumn{4}{c}{CMB-Subtracted Flux Density} \\
 & & & & \texttt{SMICA} & \texttt{NILC} & \texttt{SEVEM} & \texttt{Commander} \\
 & [GHz] & [mm] & [Jy] & \multicolumn{4}{c}{[Jy]} \\
\hline
\hline

LFI030 & 28.4 & 10.6 & 0.46 $\pm$ 0.05 & 0.55 $\pm$ 0.03 & 0.51 $\pm$ 0.03 & 0.47 $\pm$ 0.03 & 0.49 $\pm$ 0.03 \\
LFI044 & 44.1 & 6.80 & 0.61 $\pm$ 0.12 & 0.77 $\pm$ 0.08 & 0.63 $\pm$ 0.08 & 0.57 $\pm$ 0.08 & 0.64 $\pm$ 0.08 \\
LFI070 & 70.4 & 4.26 & 1.83 $\pm$ 0.12 & 1.42 $\pm$ 0.14 & 1.07 $\pm$ 0.14 & 0.79 $\pm$ 0.14 & 0.99 $\pm$ 0.14 \\
HFI100 & 100 & 3.00 & 2.93 $\pm$ 0.17 & 2.25 $\pm$ 0.09 & 1.54 $\pm$ 0.09 & 1.10 $\pm$ 0.08 & 1.45 $\pm$ 0.09 \\
HFI143 & 143 & 2.10 & 7.41 $\pm$ 0.26 & 5.99 $\pm$ 0.08 & 4.86 $\pm$ 0.08 & 4.07 $\pm$ 0.07 & 4.64 $\pm$ 0.08 \\
HFI217 & 217 & 1.38 & 21.1 $\pm$ 0.4 & 18.7 $\pm$ 0.2 & 17.2 $\pm$ 0.2 & 16.2 $\pm$ 0.2 & 17.0 $\pm$ 0.2 \\
HFI353 & 353 & 0.849 & 76.6 $\pm$ 0.8 & 75.2 $\pm$ 0.8 & 74.0 $\pm$ 0.8 & 73.1 $\pm$ 0.8 & 74.0 $\pm$ 0.8 \\
HFI545 & 545 & 0.550 & 241.0 $\pm$ 16.9 & 241.0 $\pm$ 16.9 & 239.0 $\pm$ 16.7 & 238.0 $\pm$ 16.7 & 239.0 $\pm$ 16.7 \\
SPIRE 500\,$\mu$m & 600 & 0.500 & 345.0 $\pm$ 34.5 & 344.0 $\pm$ 34.4 & 342.0 $\pm$ 34.2 & 341.0 $\pm$ 34.1 & 341.0 $\pm$ 34.1 \\
SPIRE 350\,$\mu$m & 857 & 0.350 & 768.0 $\pm$ 76.8 & 768.0 $\pm$ 76.8 & 766.0 $\pm$ 76.6 & 764.0 $\pm$ 76.5 & 766.0 $\pm$ 76.6 \\
HFI857 & 857 & 0.350 & 693.0 $\pm$ 48.6 & 694.0 $\pm$ 48.7 & 693.0 $\pm$ 48.5 & 692.0 $\pm$ 48.5 & 693.0 $\pm$ 48.6 \\
SPIRE 250\,$\mu$m & 1200 & 0.250 & 1500 $\pm$ 150 & 1500 $\pm$ 150 & 1500 $\pm$ 150 & 1500 $\pm$ 150 & 1500 $\pm$ 150 \\
PACS 160\,$\mu$m & 1870 & 0.160 & 2230 $\pm$ 334 & 2250 $\pm$ 337 & 2250 $\pm$ 338 & 2250 $\pm$ 338 & 2260 $\pm$ 338 \\
PACS 100\,$\mu$m & 3000 & 0.100 & 1400 $\pm$ 210 & 1380 $\pm$ 208 & 1380 $\pm$ 207 & 1380 $\pm$ 208 & 1380 $\pm$ 207 \\
IRIS 100\,$\mu$m & 3000 & 0.100 & 1340 $\pm$ 181 & 1330 $\pm$ 179 & 1330 $\pm$ 179 & 1330 $\pm$ 179 & 1330 $\pm$ 179 \\
PACS 70\,$\mu$m & 4280 & 0.0700 & 535.0 $\pm$ 80.3 & 531.0 $\pm$ 79.7 & 529.0 $\pm$ 79.4 & 530.0 $\pm$ 79.6 & 527.0 $\pm$ 79.1 \\
IRIS 60\,$\mu$m & 5000 & 0.0600 & 464.0 $\pm$ 48.2 & 465.0 $\pm$ 48.3 & 464.0 $\pm$ 48.2 & 468.0 $\pm$ 48.7 & 460.0 $\pm$ 47.9 \\
IRIS 25\,$\mu$m & 12000 & 0.0250 & 53.0 $\pm$ 8.0 & 51.0 $\pm$ 7.7 & 50.6 $\pm$ 7.7 & 50.0 $\pm$ 7.6 & 50.8 $\pm$ 7.7 \\
MIPS 24\,$\mu$m & 12500 & 0.0240 & 53.7 $\pm$ 5.4 & 52.6 $\pm$ 5.3 & 52.4 $\pm$ 5.2 & 52.1 $\pm$ 5.2 & 52.5 $\pm$ 5.3 \\
IRIS 12\,$\mu$m & 25000 & 0.0120 & 43.5 $\pm$ 2.3 & 43.5 $\pm$ 2.3 & 43.5 $\pm$ 2.3 & 43.5 $\pm$ 2.3 & 43.5 $\pm$ 2.3 \\
IRAC 8\,$\mu$m & 37500 & 0.00800 & 77.0 $\pm$ 7.7 & 77.0 $\pm$ 7.7 & 77.0 $\pm$ 7.7 & 77.0 $\pm$ 7.7 & 77.0 $\pm$ 7.7 \\
IRAC 5.8\,$\mu$m & 51700 & 0.00580 & 62.5 $\pm$ 6.3 & 62.5 $\pm$ 6.3 & 62.5 $\pm$ 6.3 & 62.5 $\pm$ 6.3 & 62.5 $\pm$ 6.3 \\
IRAC 4.5\,$\mu$m & 66600 & 0.00450 & 14.2 $\pm$ 1.4 & 14.2 $\pm$ 1.4 & 14.2 $\pm$ 1.4 & 14.2 $\pm$ 1.4 & 14.2 $\pm$ 1.4 \\
IRIS 3.6\,$\mu$m & 83300 & 0.00360 & 18.8 $\pm$ 1.9 & 18.8 $\pm$ 1.9 & 18.8 $\pm$ 1.9 & 18.8 $\pm$ 1.9 & 18.8 $\pm$ 1.9 \\

\hline
\label{Table:M33_Flux_Dens}
\end{tabular}
\end{center}
\vspace{-0.6cm}
\end{table*}

\subsection{Decomposition of the continuum flux density spectrum}
\label{Subsec:Decomposition}

In order to quantify the resulting flux density spectra, we fitted each of them with a model simultaneously fitting contributions representing thermal dust emission, as a combination of two modified blackbodies at different temperatures~(the use of two modified blackbodies is to insure an accurate fit to the peak of the cold dust component), but with identical dust emissivity indices, 

\begin{equation}
S_{\nu}^\mathrm{dust} = \sum\limits_{i=1}^{2} C_\mathrm{dust,i} \left( \frac{\nu}{\nu_{0}} \right)^{\beta_\mathrm{eff}} B_{\nu}(T_\mathrm{dust,i}) ,
\label{equ:S_Td}
\end{equation}

\noindent
non-thermal synchrotron emission,

\begin{equation}
S_{\nu}^\mathrm{sync} = C_\mathrm{sync} \left( \frac{\nu}{\nu_{0}} \right)^{\alpha_\mathrm{sync}} ,
\label{equ:S_sync}
\end{equation}

\noindent
free-free emission, 

\begin{equation}
S_{\nu}^\mathrm{ff} = C_\mathrm{ff} \left( \frac{\nu}{\nu_{0}} \right)^{\alpha_\mathrm{ff}} ,
\label{equ:S_ff}
\end{equation}

\noindent
and AME, assuming that it is due to spinning dust emission,

\begin{equation}
S_{\nu}^\mathrm{AME} = C_\mathrm{AME} j_{\nu} ,
\label{equ:S_AME}
\end{equation}

\noindent
where $j_{\nu}$ is the spinning dust emissivity for the warm ionised medium computed using the spinning dust model, \textsc{spdust}~\citep{Ali-Haimoud:09, Silsbee:11}. For each flux density spectrum we used the IDL fitting routine \textsc{mpfit}~\citep{Markwardt:09}, which employs the Levenberg-Marquardt least-squares minimisation technique, to fit the data between 1.4\,GHz and 24\,$\mu$m for $C_\mathrm{dust,i}$, $T_\mathrm{dust,i}$, $\beta_\mathrm{eff}$, $C_\mathrm{sync}$, $\alpha_\mathrm{sync}$, $C_\mathrm{ff}$, $\alpha_\mathrm{ff}$, and $C_\mathrm{AME}$. During the fitting process, $C_\mathrm{dust,i}$, $T_\mathrm{dust,i}$, $C_\mathrm{sync}$, $C_\mathrm{AME}$ were constrained to be physically realistic~(i.e., $\ge$0), $\beta_\mathrm{eff}$ and $\alpha_\mathrm{sync}$ were unconstrained, while $\alpha_\mathrm{ff}$ was fixed to $-$0.1 and $C_\mathrm{ff}$, which as discussed below, was constrained based on additional observations. The estimated uncertainties on these fitted parameters were computed from the resulting covariance matrix.

\begin{figure*}
\begin{center}
\includegraphics[angle=0,scale=0.75]{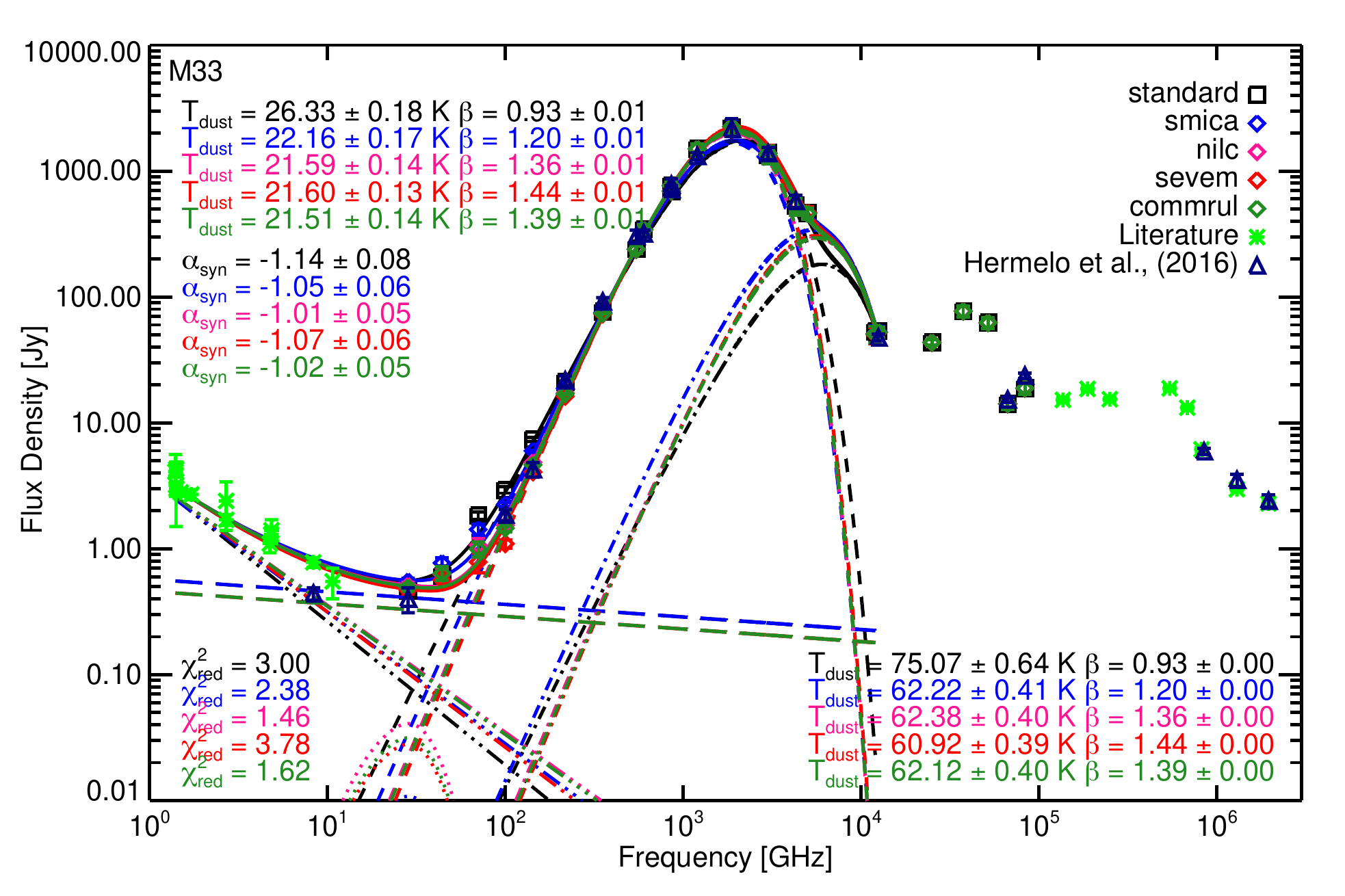}
\includegraphics[angle=0,scale=0.75]{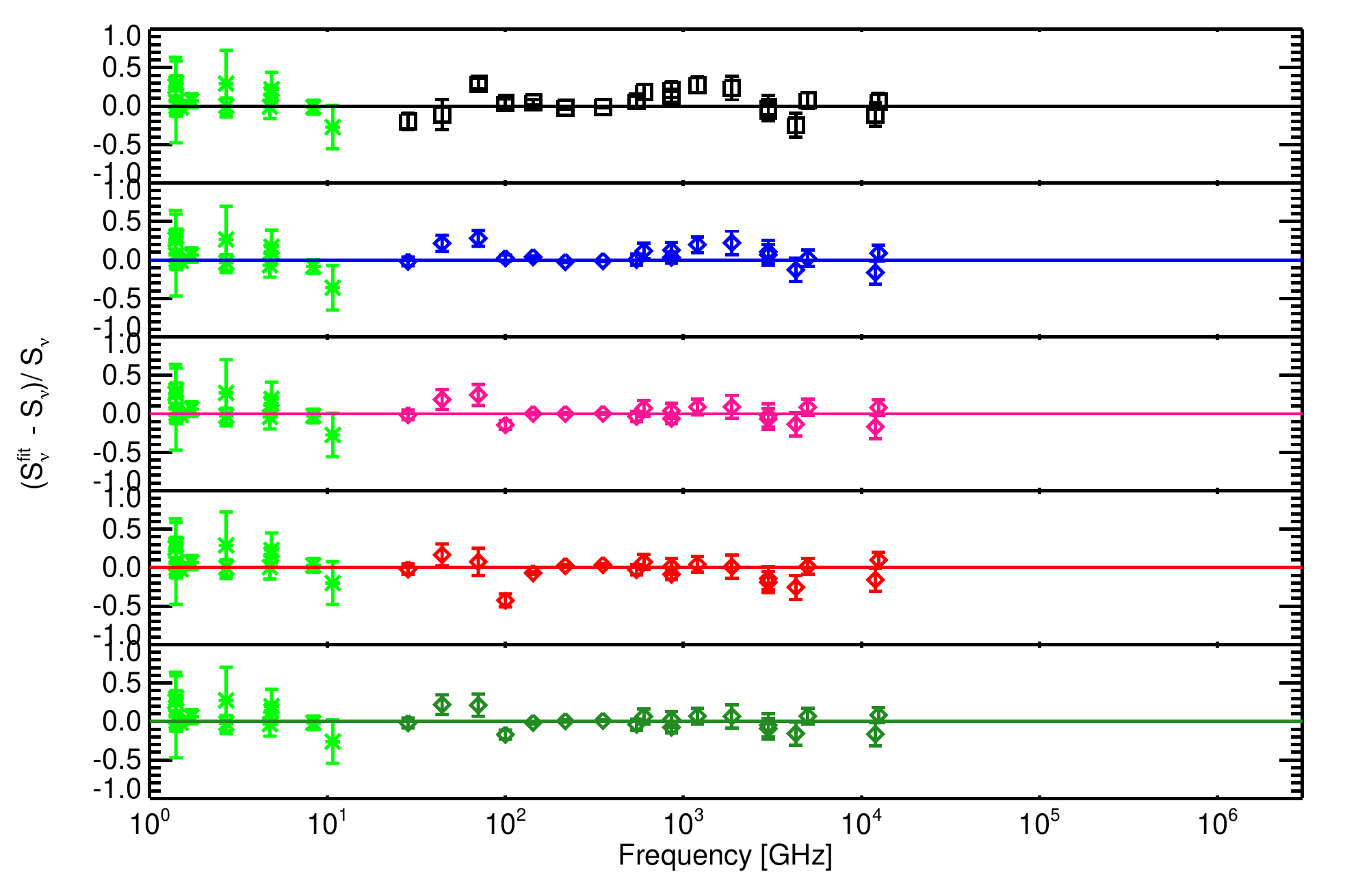}
\end{center}
\vspace{-0.3cm}
\caption{\textit{Top:} Continuum flux density spectra for M33. The fitted components include free-free emission~(long dashed line), synchrotron emission~(dot-dot-dot-dashed line), cold thermal dust emission~(dashed line), warm thermal dust emission~(dot-dashed line), and spinning dust emission~(dotted line). These components are fitted individually to the five datasets (standard \textit{Planck} data, along with the \texttt{SMICA}, \texttt{NILC}, \texttt{SEVEM}, and \texttt{Commander} CMB-subtracted data) and the resulting parameters are displayed on the plot. \textit{Bottom:} Normalised residuals of the fits to the five continuum flux density spectra for M33.}
\label{Fig:M33_Full_CFD}
\end{figure*}

\begin{table*}
\begin{center}
\caption{Best fit parameters from the fits to the M33 continuum flux density spectra displayed in Fig.~\ref{Fig:M33_Full_CFD}.}
\begin{tabular}{lccccc}
\hline
Data & $\beta_\mathrm{eff}$ & $T_\mathrm{dust}$ & $\alpha_\mathrm{sync}$ & $\alpha_\mathrm{ff}$ & $\chi^{2}_\mathrm{red}$ \\
 
 & & [K] \\
\hline
\hline

standard		 		&	0.93~$\pm$~0.01 &		26.33~$\pm$~0.18 & 	$-$1.14~$\pm$~0.08 &		$-$0.1~(fixed) &	3.00      \\
\texttt{SMICA}			&	1.20~$\pm$~0.01 &		22.16~$\pm$~0.17 & 	$-$1.05~$\pm$~0.06 &		$-$0.1~(fixed) &	2.38	      \\
\texttt{NILC} 			& 	1.36~$\pm$~0.01 &		21.59~$\pm$~0.14 & 	$-$1.01~$\pm$~0.05 &		$-$0.1~(fixed) &	1.46	      \\
\texttt{SEVEM}			&	1.44~$\pm$~0.01 &		21.60~$\pm$~0.13 & 	$-$1.07~$\pm$~0.06 &		$-$0.1~(fixed) &	3.78	      \\
\texttt{Commander}		&	1.39~$\pm$~0.01 &		21.51~$\pm$~0.14 & 	$-$1.02~$\pm$~0.05 &		$-$0.1~(fixed) &	1.62	      \\
\\
Mean of CMB-subtracted 	&	1.35~$\pm$~0.10 &		21.67~$\pm$~0.30 & 	$-$1.03~$\pm$~0.03 &		$-$0.1~(fixed) &	-	      \\

\hline
\label{Table:Fitted_Parameters}
\end{tabular}
\end{center}
\vspace{-0.6cm}
%Table notes here
\end{table*}

\begin{figure}
\begin{center}
\includegraphics[angle=0,scale=0.44]{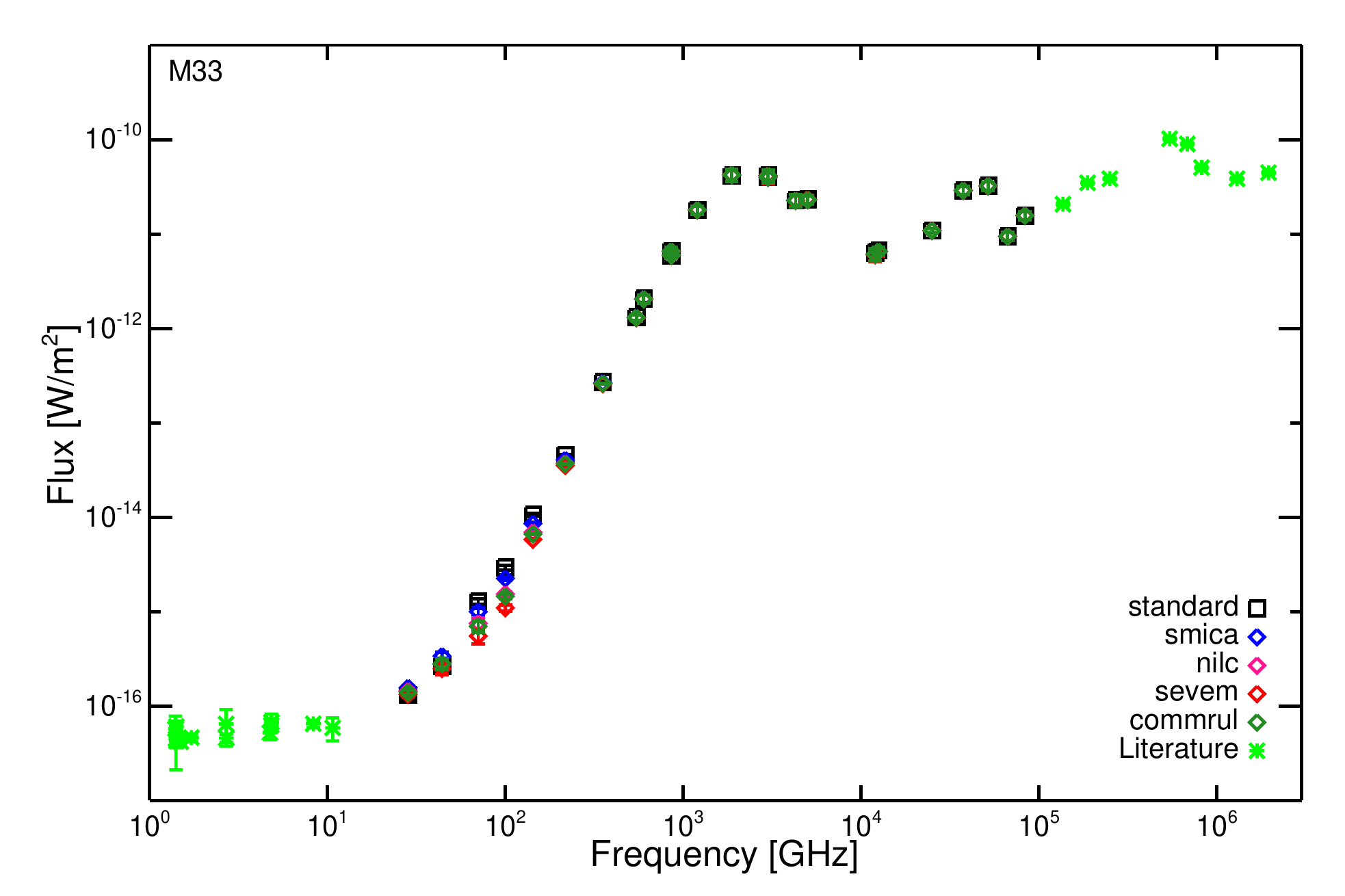}
\end{center}
\vspace{-0.3cm}
\caption{SEDs for M33 for the five datasets (standard \textit{Planck} data, along with the \texttt{SMICA}, \texttt{NILC}, \texttt{SEVEM}, and \texttt{Commander} CMB-subtracted data).}
\label{Fig:M33_Full_SED}
\end{figure}

In order to derive reliable thermal dust parameters, the unrelated contributions of the thermal~(free-free) and non-thermal~(synchrotron) emission components of the gas must be accurately determined. Unfortunately, the decomposition of the low-frequency radio continuum of galaxy flux density spectra is usually not clear-cut because of the degeneracy between the free-free and synchrotron contributions, especially since the intrinsic spectral index of the synchrotron emission is not known. This degeneracy specifically hampers the determination of any AME contribution to the observed emission spectrum and additional constraints are desirable. In the case of M33, such constraints exist. The sum of the directly measured H\textsc{ii} region flux densities corresponds to 235\,mJy at 10\,GHz~\citep{Israel:74, Israel:80}. Unfortunately, this does not include any contribution from the diffuse emission and the tally of H\textsc{ii} regions is incomplete, especially at large radii. It thus provides us only with a useful lower limit. However, a more accurate estimate of the total thermal radio emission may be obtained from the integrated H$\alpha$ emission~($I_\mathrm{H\alpha}$ = 3.6$\times$10$^{-13}$\,W\,m$^{-2}$) measured by~\citet{Hoopes:00}, after first correcting for global extinction. The M33 SED shown in Fig.~\ref{Fig:M33_Full_SED} exhibits a peak at optical wavelengths~($\sim$5$\times$10$^{5}$\,GHz) representing the directly observed integrated starlight, and another peak in the far-IR~($\sim$2$\times$10$^{3}$\,GHz) representing absorbed and re-emitted starlight. Since the second peak is less than the first one, a relatively minor fraction of all starlight is intercepted by dust, and the (small) global extinction can be estimated from the ratio of the peak fluxes. The M33 SED resembles those of the SMC and, in particular, the LMC~\citep{Israel:10}, with an optical luminosity exceeding the IR luminosity by a factor of~$\sim$2.5. From the luminosity ratio of the far-IR and optical peaks in Fig.~\ref{Fig:M33_Full_SED}, we estimate a visual extinction A$_{V}$~=~0.4~$\pm$~0.1\,mag, of which 0.1\,mag is due to the Milky Way foreground~\citep{DeVaucouleurs:91}. Assuming A$_\mathrm{H\alpha}$ = 0.81A$_{V}$, which is a typical Milky Way extinction curve~\citep{Fitzpatrick:07}, and at these wavelengths is very similar to typical SMC and LMC extinction curves~\citep{Gordon:03}, this corresponds to an H$\alpha$ extinction A$_\mathrm{H\alpha}$ = 0.24~$\pm$~0.08\,mag \textit{internal} to M33. Hence, the corrected H$\alpha$ flux is (4.5~$\pm$~0.5) $\times$ 10$^{-13}$\,W\,m$^{-2}$. Using

\begin{equation}
\frac{S^\mathrm{ff}_{\nu}}{I_\mathrm{H\alpha}} = 1.15\times10^{-14} [1-0.21\times\mathrm{log}\left(\frac{\nu}{\mathrm{GHz}}\right)]~\mathrm{Hz^{-1}} ,
\label{equ:S_Halpha}
\end{equation} 

\noindent 
we find that the free-free flux density at 10\,GHz is $S^\mathrm{ff}_\mathrm{10\,GHz}$ = 410~$\pm$~45\,mJy. This is higher than the value of 280\,mJy inferred from the work by~\citet{Buczilowski:88}, but in agreement with the value of 400\,mJy that follows from the determination by~\citet{Tabatabaei:07a, Tabatabaei:07b}. Therefore, during the fitting process we constrained $C_\mathrm{ff}$ to be 410~$\pm$~45\,mJy at 10\,GHz, and fixed $\alpha_\mathrm{ff}$~=~$-$0.1. This is an important element in the decomposition of the observed continuum spectrum, essential for a reliable evaluation of the AME contribution in the 5--50\,GHz frequency range.

The full results of our fitting analysis are shown in Fig.~\ref{Fig:M33_Full_CFD}, including the fitted parameters~(which are also listed in Table~\ref{Table:Fitted_Parameters}), the individual fitted components, the overall flux density spectrum fit, and the normalise residuals of the fits. It is clear that there is significant difference between the fit to the standard \textit{Planck} data compared to the CMB-subtracted \textit{Planck} data. Although there is a spread in the fitted $T_\mathrm{dust}$ and $\beta_\mathrm{eff}$ values estimated from the CMB-subtracted data~(blue, pink, red, and green curves in Fig.~\ref{Fig:M33_Full_CFD}), the fit to the standard data~(black curve in Fig.~\ref{Fig:M33_Full_CFD}) is significantly outside this range. Focusing solely on the fits to the CMB-subtracted data, and combing these four fits, we find that the dust emission spectrum of M33 between~$\sim$100\,GHz and 3\,THz is adequately described by a single modified blackbody curve, with a peak of~$\sim$2000\,Jy, a mean dust temperature $T_\mathrm{dust}$ = 21.67~$\pm$~0.30\,K, and a mean effective dust emissivity $\beta_\mathrm{eff}$ = 1.35~$\pm$~0.10. There is also a warm dust component with a mean temperature of 61.89~$\pm$~0.67\,K that was forced to have the same effective dust emissivity as the cold dust component. Since this warm component was simply included to insure an accurate fit to the peak of the cold dust component, we do not interpret this any further. The mean synchrotron radio continuum spectral index is $\alpha_\mathrm{sync}$ = $-$1.03~$\pm$~0.03. The relevant mean values are also listed in Table~\ref{Table:Fitted_Parameters}. For the individual entries we list the internal errors, whereas the errors given for the mean values reflect the rather larger dispersion of the individual values. Comparing the mean values to the standard values (i.e., comparing the first and last rows in Table~\ref{Table:Fitted_Parameters}) we find that not correcting for the CMB contribution would result in a significant over-estimate of $T_\mathrm{dust}$ (by~$\sim$5\,K) and under-estimate of $\beta_\mathrm{eff}$ (by~$\sim$0.4), clearly highlighting the importance of correcting for the CMB. 

We computed the mean fraction of thermal radio emission at 20\,cm and 3.6\,cm to be~$\sim$16\,\% and~$\sim$49\,\%, respectively, which are consistent with the estimates from~\citet{Tabatabaei:07b}, confirming that our estimate of the thermal emission from the H$\alpha$ emission is reasonable. Not surprisingly, our estimated synchrotron spectral index is also consistent with the results obtained by~\citet{Tabatabaei:07b}. Although our fitted free-free emission amplitudes match the limits of the estimate obtained from the H$\alpha$ observations, we confirmed that even without imposing this constraint on the free-free emission, we find consistent results, with a mean dust temperature of $T_\mathrm{dust}$~=~22.36~$\pm$~0.69\,K and a mean spectral index of $\beta_\mathrm{eff}$~=~1.31~$\pm$~0.10, but the fraction of free-free emission is decreased to~$\sim$11\,\% and~$\sim$34\,\% at 20\,cm and 3.6\,cm, respectively. Ignoring the radio data, and simply fitting the data at frequencies $\ge$\,100\,GHz, we find a slightly lower value of $\beta_\mathrm{eff}$ = 1.28~$\pm$~0.10. The fact that these different approaches all yield consistent results confirms that our fit is not biased by the radio data, the decomposition between the thermal and non-thermal radio emission, nor the degeneracy between the free-free emission and the AME.

We find that the AME in M33 is at best a minor component, both in an absolute sense and when compared to the free-free and synchrotron emission. Even though we modelled the AME using a spinning dust model for the warm ionised medium~(as was used by \citet{Planck_Intermediate_Results_XXV:15} for their M31 analysis), we obtained consistent results using the cold neutral medium, warm neutral medium, or molecular cloud spinning dust models. Based on observations of AME in the Milky Way, the ratio between the AME emission at 30\,GHz and the thermal dust emission at 100\,$\mu$m~(often incorrectly referred to as an AME emissivity) is of the order of~$\sim$2$\times$10$^{-4}$~\citep{Todorovic:10, Planck_Intermediate_Results_XV:14}. Therefore, since we find a 100\,$\mu$m flux density of~$\sim$1350\,Jy, this would lead us to expect~$\sim$0.3\,Jy of AME at 30\,GHz, while we only estimate an AME flux density of $\lesssim$\,0.04\,Jy. Although this ratio between the AME and thermal dust emission is sensitive to dust temperature~\citep[as discussed by][]{Tibbs:12b}, these results indicate that there is significantly less AME in M33 compared to our own Galaxy, which is consistent with what has been observed in M82, NGC253, and NGC4945~\citep{Peel:11}. On the other hand, in M31 the AME appears to be much more prominent, with a tentative detection that is comparable to what would be expected based on the AME level observed in our own Galaxy~\citep{Planck_Intermediate_Results_XXV:15}. However, we note that for M31, the lack of observations between~$\sim$1 and 20\,GHz could bias the fit, which is not the case for M33, M82, NGC253, and NGC4945, where the wavelength coverage is more complete.

Finally, we emphasise that a single dust temperature and a single effective emissivity index, i.e., a single curve, provides a reasonable fit to the observed data between~$\sim$100\,GHz and 3\,THz. There is no spectral break, and there is no indication of an ``excess'' of any kind in the global (spatially integrated) flux density spectrum of M33.

\begin{figure}
\begin{center}
\includegraphics[angle=0,scale=0.4555]{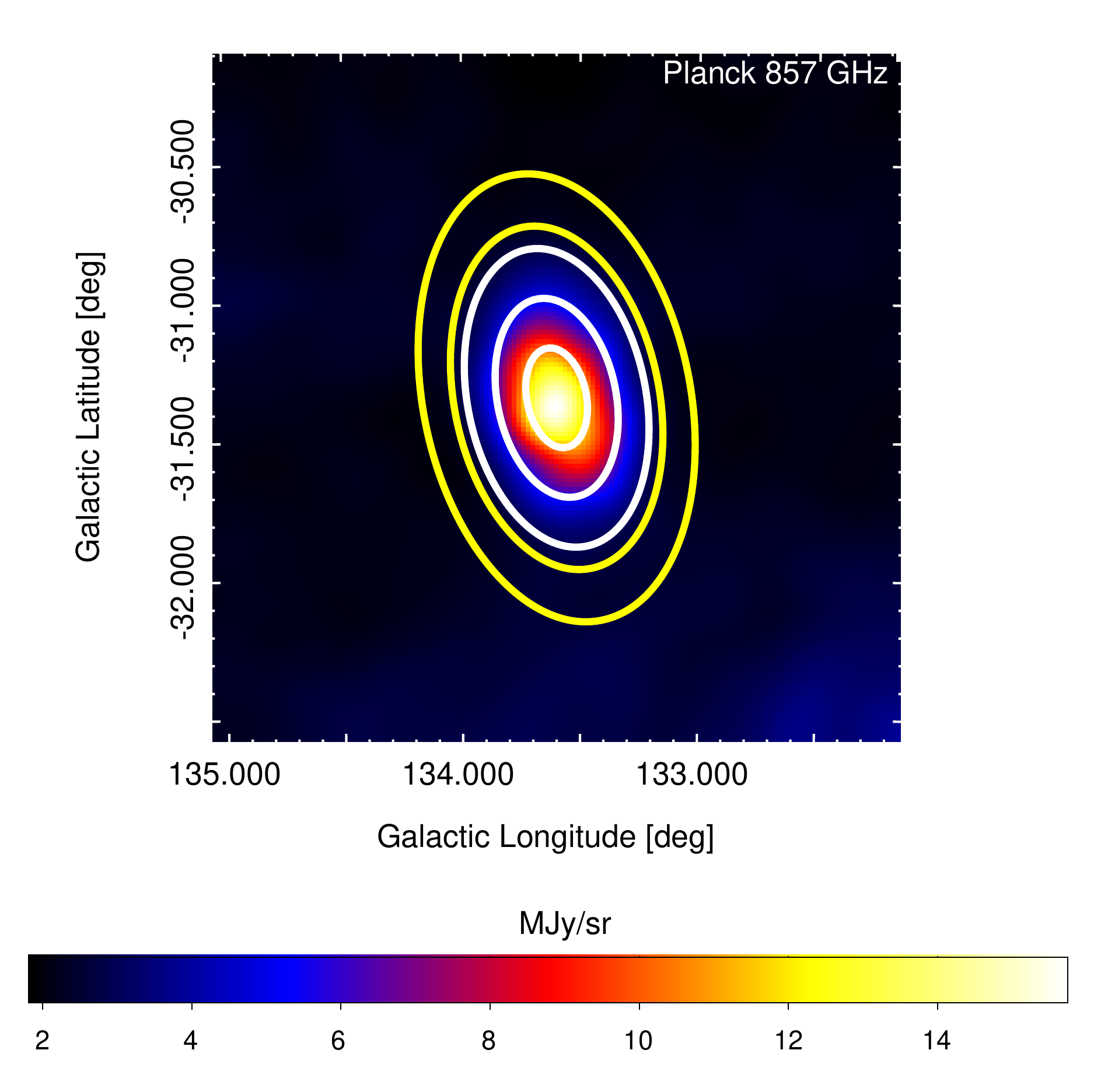}
\end{center}
\vspace{-0.3cm}
\caption{\textit{Planck} 857\,GHz map convolved to 10\,arcmin angular resolution. Superimposed are the three independent elliptical apertures (white ellipses with semi-major axes, $a$~=~8, 5.33, and 2.67\,kpc) and the background annulus (green ellipses). The semi-major to semi-minor axis ratio was fixed to 10$^{0.23}$.}
\label{Fig:M33_PLCK857_10arcmin}
\end{figure}

\begin{figure*}
\begin{center}
\includegraphics[angle=0,scale=0.44]{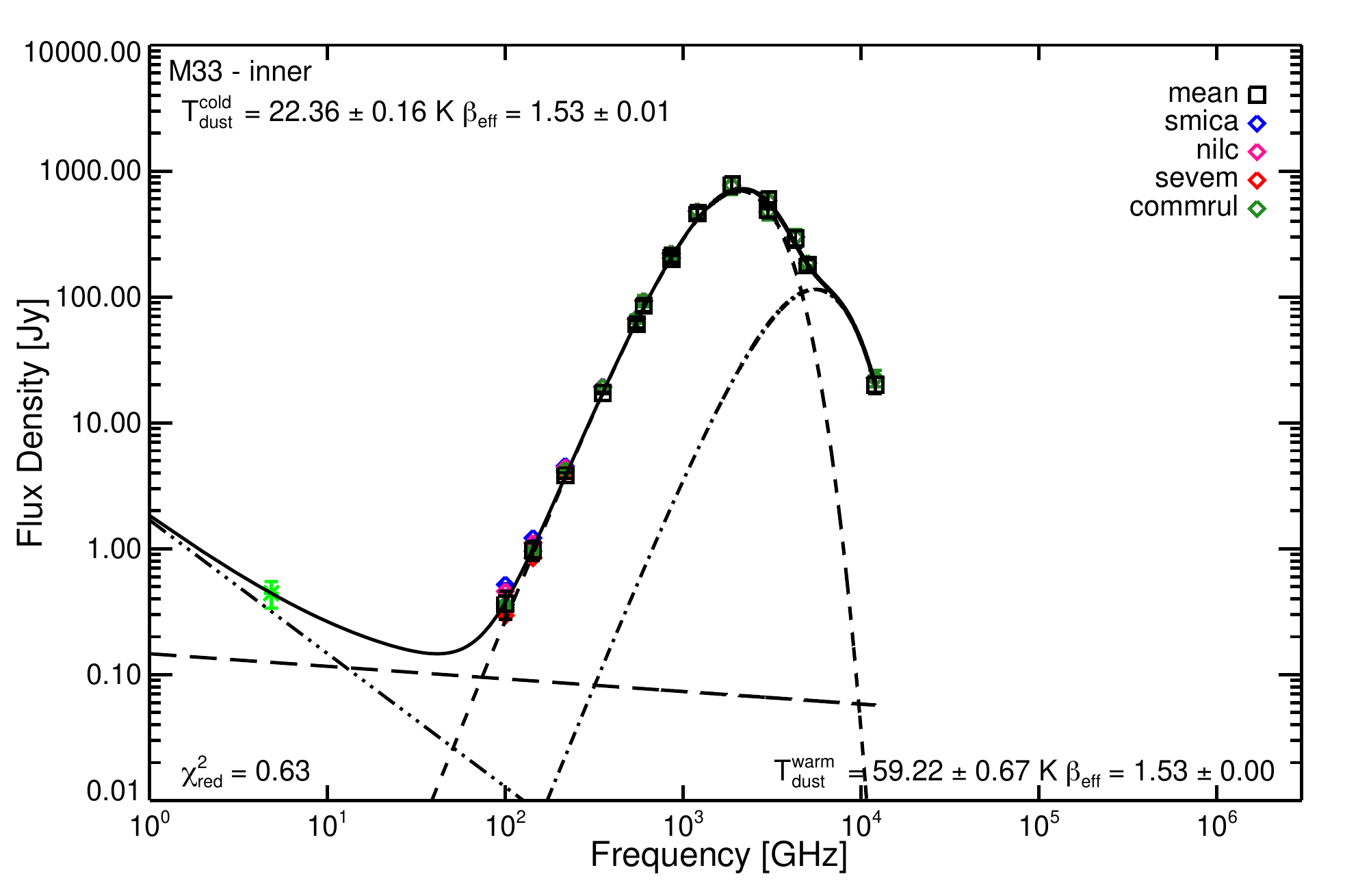}
\includegraphics[angle=0,scale=0.44]{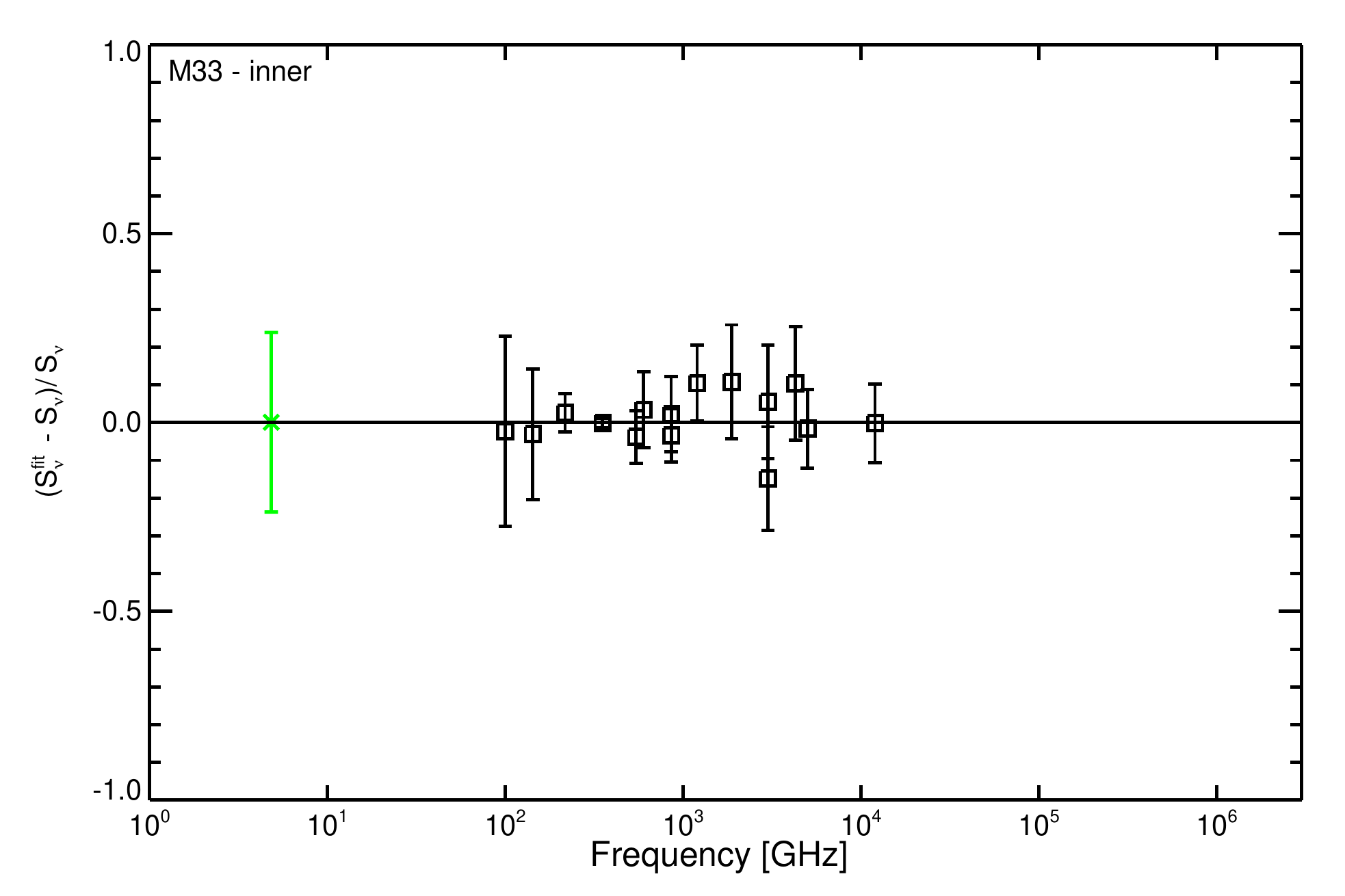} \\
\includegraphics[angle=0,scale=0.44]{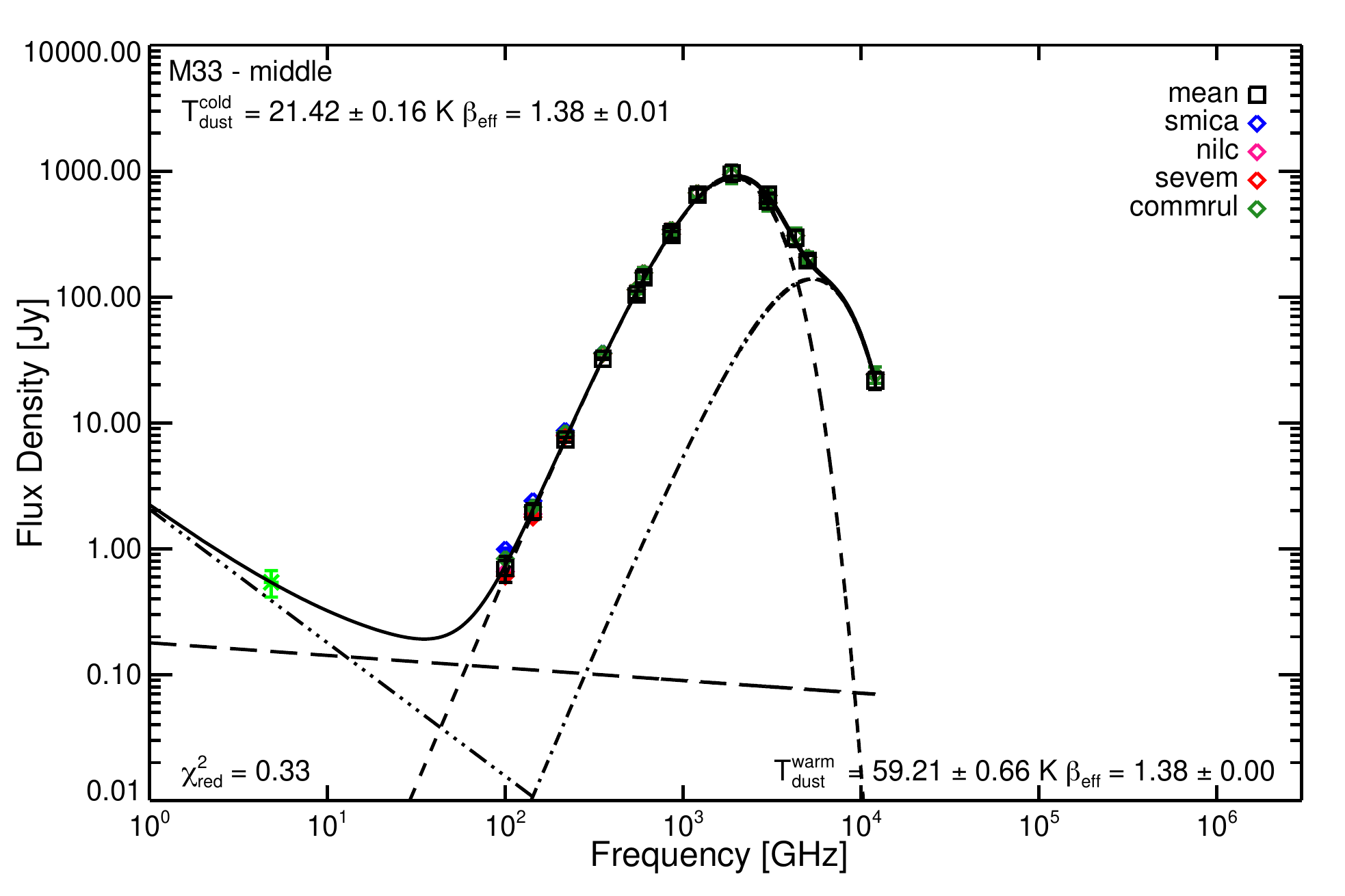}
\includegraphics[angle=0,scale=0.44]{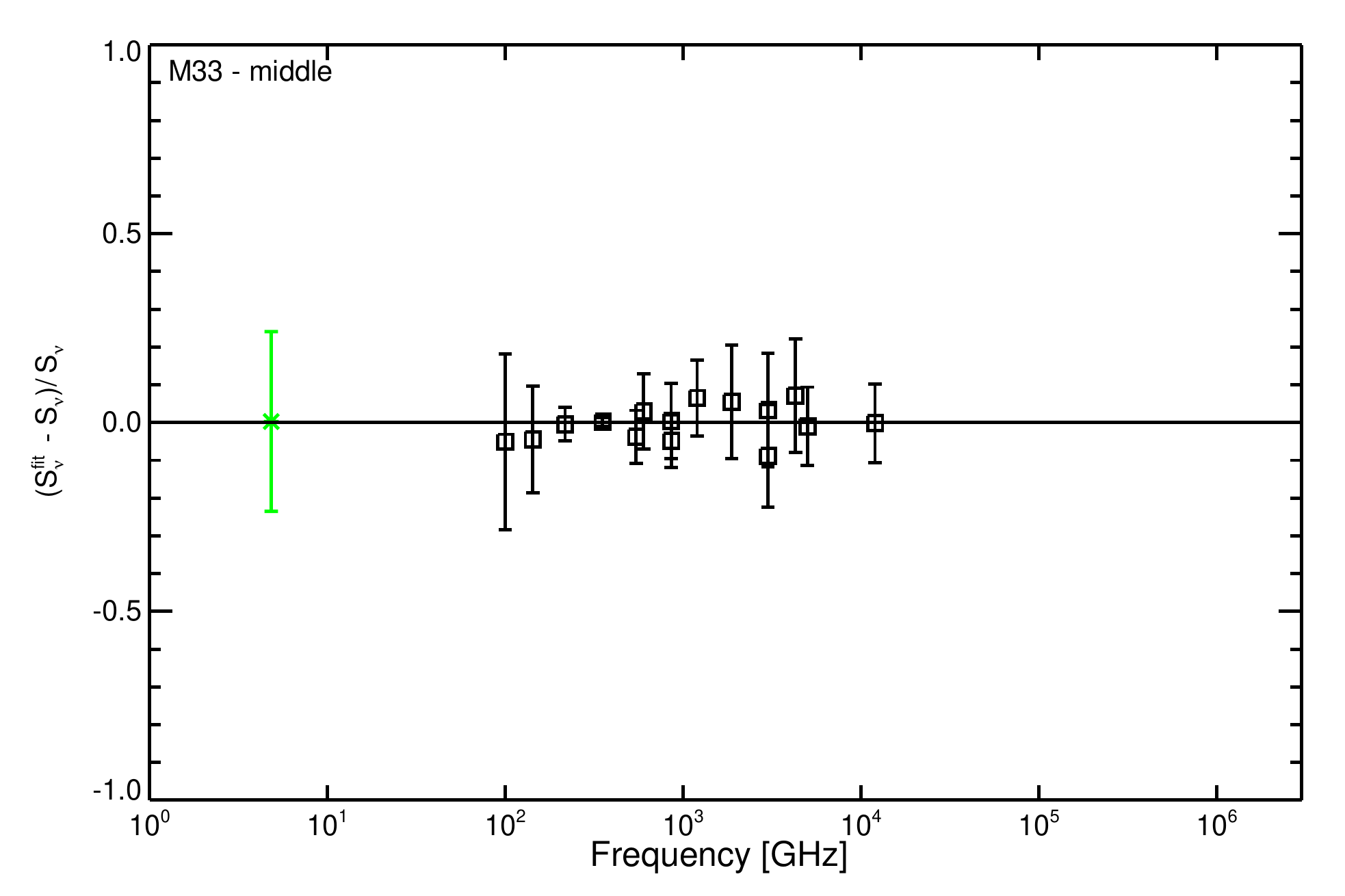} \\
\includegraphics[angle=0,scale=0.44]{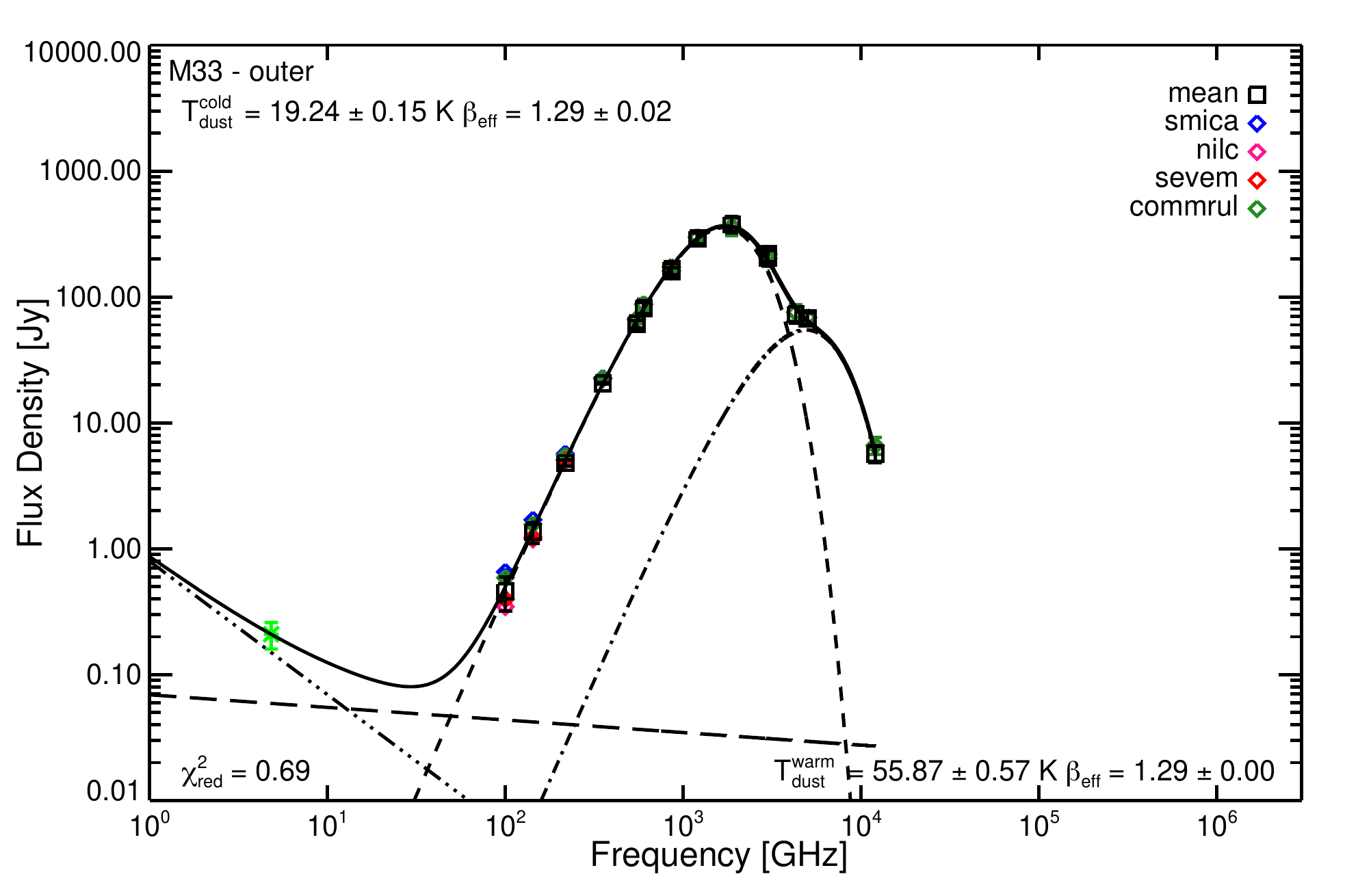}
\includegraphics[angle=0,scale=0.44]{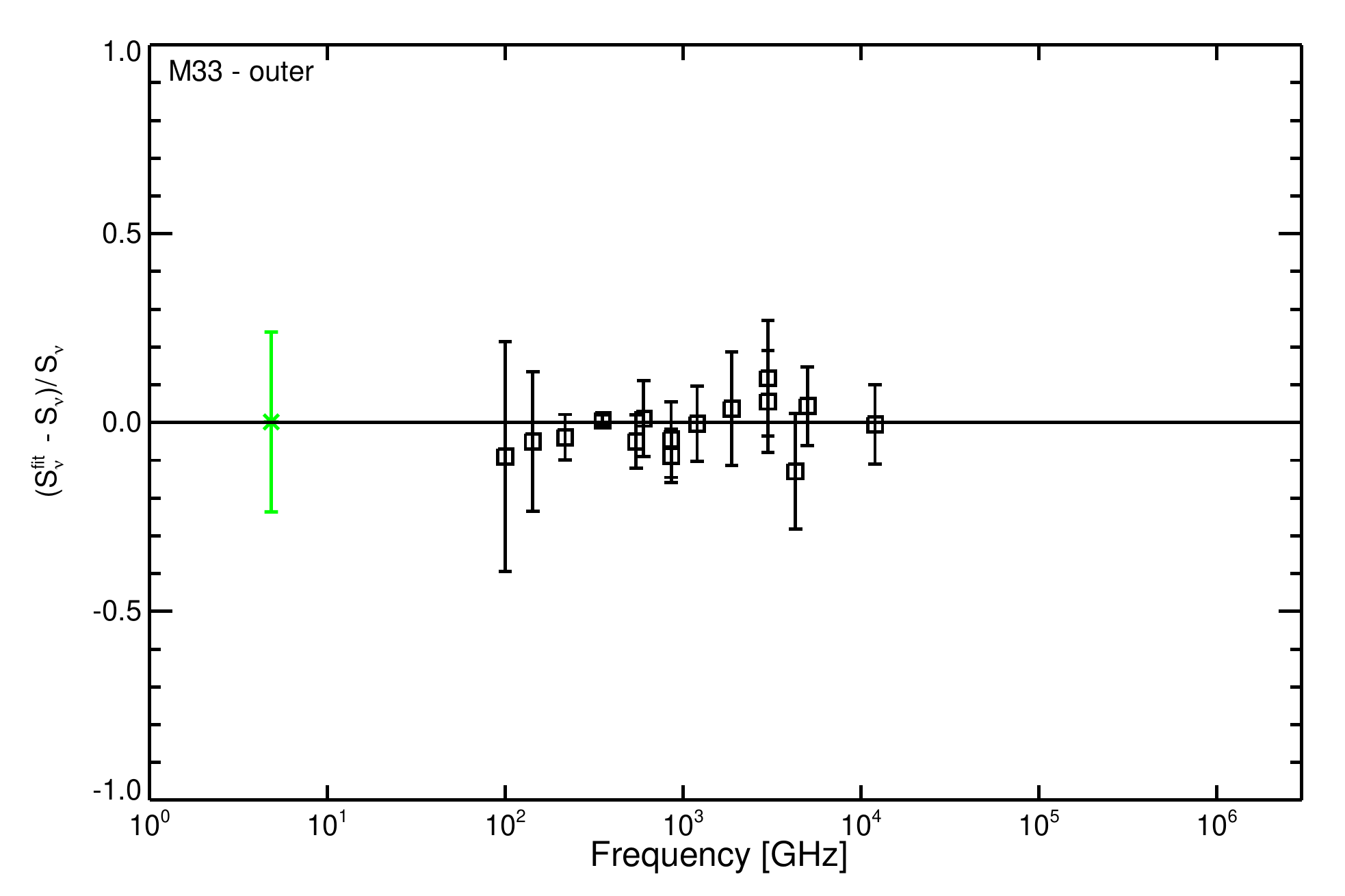} \\
\end{center}
\vspace{-0.3cm}
\caption{Continuum flux density spectra~(\textit{left}) and the corresponding normalised residuals for the fit~(\textit{right}) for the inner~(\textit{top}), middle~(\textit{middle}), and outer~(\textit{bottom}) regions of M33. The fitted components include free-free emission~(long dashed line), synchrotron emission~(dot-dot-dot-dashed line), cold thermal dust emission~(dashed line), warm thermal dust emission~(dot-dashed line), and spinning dust emission~(dotted line -- not shown). The resulting parameters of the fits are displayed on each spectrum.}
\label{Fig:M33_radial}
\end{figure*}

\subsection{Radial variations}
\label{Subsec:Annuli}

In view of the radial changes in the brightness of the M33 disk, it is of interest to establish whether or not the flux density spectrum changes with radial distance from the center of M33. For this purpose, we now use the \textit{Planck}, \textit{Herschel}, and \textit{IRAS} maps at the best common angular resolution (10\,arcmin), ignoring the three \textit{Planck} LFI bands, which have beams larger than 10\,arcmin. Even so, the limited extent of M33 allows only three fully independent concentric ellipses to be constructed with semi-major axes of 8, 5.33, and 2.67\,kpc. As before, the background/foreground emission was estimated within an elliptical annulus with inner and outer semi-major axes of 1.15 and 1.50 times the semi-major axis of the 8\,kpc aperture, respectively~(see Fig.~\ref{Fig:M33_PLCK857_10arcmin}).

\citet{Tabatabaei:07a, Tabatabaei:07b} have shown that across the disk of M33, the radial profiles of the far-IR and radio emission are very similar. Therefore, in addition to fitting the far-IR/sub-mm emission within each annulus, we also approximated the radio emission by scaling the curves from Fig.~\ref{Fig:M33_Full_CFD}. To do this, we computed the mean ratio between the 160\,$\mu$m flux density and the 4.8\,GHz flux density from the analysis in Section~\ref{Subsec:Full_SED}, and assuming that this is constant, we estimated the level of the radio emission within each annulus. We also assume that the fraction of free-free emission at 4.8\,GHz is fixed across M33, and fit the total radio emission with a synchrotron spectral index identical to the mean found for the entire galaxy. As before, we fit for synchrotron, free-free, AME, and thermal dust emissions. However, for this analysis we combine the four flux density measurements at each wavelength using the scatter as a measure of the uncertainty and perform a single fit, rather than a separate fit for each of the four CMB-subtracted maps. The resulting flux density spectra for each of the three elliptical annuli representing the inner, middle, and outer regions of M33 are displayed in Fig.~\ref{Fig:M33_radial}, along with the corresponding normalised residuals of the fits.

The radial dependence of both the dust temperature, $T_\mathrm{dust}$, and the effective emissivity, $\beta_\mathrm{eff}$, can be inferred from the plots, and it is clear that \textit{both} dust temperature and the effective dust emissivity decrease with increasing radius. In the center of M33, $T_\mathrm{dust}$~=~22.36~$\pm$~0.16\,K and $\beta_\mathrm{eff}$~=~1.53~$\pm$~0.01. As the radius increases, both decrease to $T_\mathrm{dust}$~=~21.42~$\pm$~0.16\,K and $\beta_\mathrm{eff}$~=~1.38~$\pm$~0.01 in the middle annulus~(semi-major axis between 2.67 and 5.33\,kpc) and to $T_\mathrm{dust}$~=~19.24~$\pm$~0.15\,K and $\beta_\mathrm{eff}$~=~1.29~$\pm$~0.02 in the outer annulus~(semi-major axis between 5.33 and 8.00\,kpc). We emphasize that this result cannot be caused by the $T_\mathrm{dust}$, $\beta_\mathrm{eff}$ degeneracy referred to earlier as that would require $\beta_\mathrm{eff}$ to increase as $T_\mathrm{dust}$ decreases.

Our finding that both $T_\mathrm{dust}$ and $\beta_\mathrm{eff}$ decrease with radius is qualitatively similar to the result obtained by~\citet{Tabatabaei:14}, who used Monte Carlo simulations on more limited \textit{Herschel} flux density spectra between 100 and 500\,$\mu$m to deduce a simultaneous decrease of dust temperature (from~$\sim$24\,K to~$\sim$18\,K) and emissivity (from~$\sim$1.8 to~$\sim$1.2) going from the center of M33 out to a radius of 6\,kpc. The more extended spectral coverage presented here allow quantitatively more robust results even though the spatial resolution is lower.

%%%%%%% Discussion %%%%%%%%%%%%%%%%%%%%%%%%%%%%%%%%%%%%%%%%%%%%%%%%

\section{Discussion}
\label{Sec:Discussion}

The \textit{global} continuum flux density spectrum of M33 is characterised by an overall emissivity $\beta_\mathrm{eff}$ = 1.35~$\pm$~0.10, which is below the value of 1.5 estimated from combined \textit{Herschel} and \textit{Spitzer} observations down to 600\,GHz (500\,$\mu$m) by~\citet{Xilouris:12}. The difference illustrates the bias introduced by the lack of low-frequency flux densities that most tightly constrain the Rayleigh-Jeans slope of the flux density spectrum. Even though we can fit the Rayleigh-Jeans part of the M33 global flux density spectrum with a single modified blackbody, it is a priori not likely that all of the dust in M33 radiates at a single temperature. However, a superposition of modified blackbodies representing grains with emissivities, $\beta_\mathrm{g}$, radiating at a range of temperatures may create a profile that is observationally hard to distinguish from a single-temperature modified blackbody profile with an apparent emissivity $\beta_\mathrm{eff}\leq\beta_\mathrm{g}$, especially when dust temperature and emissivity are negatively correlated as originally suggested by~\citet{Dupac:03} and~\citet{Desert:08}, and later confirmed by~\citet{Planck_2013_Results_XI:14}. Our analysis clearly shows that this is the case. The global flux density spectrum, whose Rayleigh-Jeans part is well-defined by a \textit{single} modified blackbody is shown to be the sum of at least three different flux density spectra representing the inner, middle, and outer regions of M33, each with Rayleigh-Jeans sections equally well fitted by a single modified blackbody. As the number of sub-spectra is only limited by the available angular resolution, we expect that each of these in turn could be decomposed further.

\subsection{Dust mass}
\label{Subsec:Mass}

Using

\begin{equation}
M_\mathrm{dust} = \frac{S_{\nu} d^{2}}{\kappa_{\nu} B_{\nu}(T_\mathrm{dust})} ,
\label{equ:Dust_Mass}
\end{equation} 

\noindent
where $\kappa_{\nu}$ is the dust opacity, we estimated the global dust mass of M33, along with the dust mass in each of the three annuli. It is known that values of the dust opacity in the literature can vary by orders of magnitude~\citep[see][]{Clark:16}, and in this work we adopt a value of $\kappa_{\nu}$ = 1.4\,m$^{2}$kg$^{-1}$ at 160\,$\mu$m taken from the ``standard model'' dust properties from~\citet{Galliano:11}. Incorporating our results from Sections~\ref{Subsec:Decomposition} and~\ref{Subsec:Annuli} into Equation~\ref{equ:Dust_Mass}, we estimated a global dust mass for M33 of (2.3$\pm$0.4$)\times$10$^{6}$ M$_{\odot}$, and (0.8$\pm$0.1)$\times$10$^{6}$ M$_{\odot}$, (1.1$\pm$0.2)$\times$10$^{6}$ M$_{\odot}$, and (0.7$\pm$0.1)$\times$10$^{6}$ M$_{\odot}$ for the inner, middle, and outer regions of M33, respectively. We find that our global dust mass estimated assuming a single modified blackbody is consistent, within the uncertainties, with the sum of the three dust masses estimated for the sub-regions, suggesting that fitting the entirety of M33 is degenerate with fitting the three sub-regions.

\subsection{Local Group sample}
\label{Subsec:Local_Group}

\begin{figure}
\begin{center}
\includegraphics[angle=0,scale=0.495]{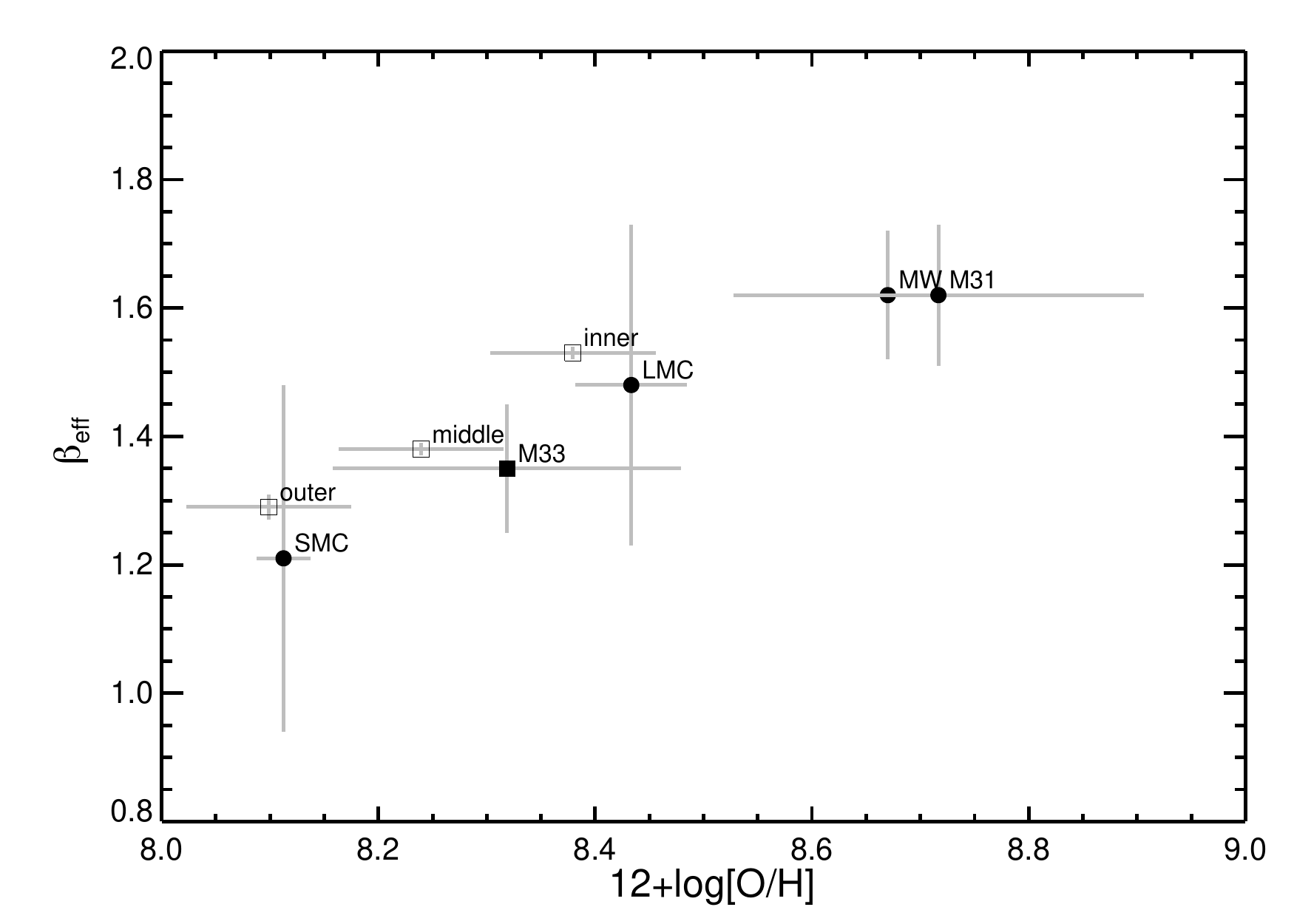}
\end{center}
\vspace{-0.3cm}
\caption{The global effective emissivity index, $\beta_\mathrm{eff}$, of Local Group galaxies as a function of metallicity, 12+log[O/H]. Filled symbols mark the global values of the identified galaxies, while the open squares mark the three independent regions (inner, middle, and outer) within M33.}
\label{Fig:M33_beta_z}
\end{figure}

The global dust emissivity, $\beta_\mathrm{eff}$ = 1.35~$\pm$~0.10, for M33 may also be compared to those derived from \textit{Planck} observations of other Local Group galaxies: $\beta_{eff}$ = 1.62~$\pm$~0.10, 1.62~$\pm$~0.11, 1.48~$\pm$~0.25, and 1.21~$\pm$~0.27 for the Milky Way~\citep{Planck_2013_Results_XI:14}, M31~\citep{Planck_Intermediate_Results_XXV:15}, the LMC~\citep{Planck_Early_Results_XVII:11}, and the SMC~\citep{Planck_Early_Results_XVII:11}, respectively. We find that the M33 emissivity is significantly lower than that observed in the Milky Way and M31, and is more consistent with the values found in the Magellanic Clouds, with M33 actually falling between the LMC and the SMC values. Interestingly, these dust emissivities closely follow the mean metallicities of the Local Group galaxies: 12+log[O/H] = 8.32~$\pm$~0.16, 8.67~$\pm$~0.04, 8.72~$\pm$~0.19, 8.43~$\pm$~0.05, and 8.11~$\pm$~0.03, for M33, the Milky Way, M31, the LMC, and the SMC, respectively~\citep[][]{Pagel:03, Toribio:16}. To illustrate this, in Fig.~\ref{Fig:M33_beta_z} we plot the dust emissivity as a function of metallicity for these 5 galaxies (filled symbols), which clearly shows that the dust emissivity increases with increasing metallicity. This trend is also observed across the M33 disk itself, where our observed emissivity gradient follows the metallicity gradient~\citep[][]{Toribio:16}, as can be seen when we plot the results from our three annuli within M33~(open squares) in Fig.~\ref{Fig:M33_beta_z}.

\subsection{$T_\mathrm{dust}$ and $\beta_\mathrm{eff}$ radial variations}
\label{Subsec:Radial_Variations}

The apparent decrease in both $T_\mathrm{dust}$ and $\beta_\mathrm{eff}$ with increasing M33 radius was discussed in some detail by~\citet{Tabatabaei:14}. Without fully subscribing to their conclusion, we note that there are, in principle, two possible physical explanations for the observed radial decreases. The first involves dust grain composition and di-electric properties. For instance, the dust emissivity may decrease with the average interstellar energy density. Mechanical and radiative erosion of dust grains should be stronger in the more energetic inner regions than in the more quiescent outer regions. This would favour more delicate carbon/ice dust grains in the outer regions and more robust silicate-rich grains in the inner regions. The intrinsic dust grain composition may also undergo radial changes following radial gradients in the population of stellar dust producers. The second explanation involves large-scale dust cloud properties. Dust cloud heating and effective emissivity may decrease with the average radiation field, more specifically the mix of dust cloud temperatures within a specific temperature range many change as a function of irradiation. For instance, consider the possibility that each of the profiles in Fig.~\ref{Fig:M33_radial} actually represent a collection of \textit{dust clouds and filaments} with identical emissivities but different temperatures. In a radially decreasing \textit{average} radiation field, clouds with temperatures at the high end would occur less frequently and have a smaller filling factor, resulting in a more skewed composite profile with a downward shift of apparent mean temperature and a consequent flattening of the Rayleigh-Jeans slope. However, the results presented in this analysis do not allow us to distinguish between these possibilities.

\subsection{Comparison to previous studies}
\label{Subsec:Hermelo}

Our analysis is not the first to produce a full flux density spectrum of M33, as \citet{Hermelo:16} used the \textit{Planck} 2013 ``nominal'' mission data along with a single CMB-subtraction method~(\texttt{SMICA}) to derive the full flux density spectrum for M33. Using complex models~\citep{Groves:08, Popescu:11}, they fitted their M33 flux density spectrum, deriving an excess of emission at mm/sub-mm wavelengths. However, in this analysis, we not only use the most recent \textit{Planck} 2015 ``full'' mission data, but we also incorporate and evaluate four different CMB-subtraction techniques. As we have discussed in some depth, the contribution from the CMB fluctuations is significant and must be accurately accounted for. From Fig.~\ref{Fig:M33_Full_CFD} and Table~\ref{Table:Fitted_Parameters} it is clear that there is scatter in both $T_\mathrm{dust}$ and $\beta_\mathrm{eff}$ between each of the four CMB-subtraction methods, highlighting the dangers of adopting a single method. 

In this analysis, we have chosen to fit our M33 flux density spectrum with a relatively simple model (i.e., a modified blackbody) and find no indication of any emission excess. A major advantage of fitting a modified blackbody to the data, rather than a complex model, is that this approach has been adopted by previous \textit{Planck} analyses~\citep[e.g.,][]{Planck_Early_Results_XVII:11, Planck_2013_Results_XI:14, Planck_Intermediate_Results_XXV:15}, allowing direct comparisons to be made with other galaxies.

%%%%%%% Conclusions %%%%%%%%%%%%%%%%%%%%%%%%%%%%%%%%%%%%%%%%%%%%%%%%

\section{Conclusions}
\label{Sec:Conclusions}

We have performed a comprehensive analysis of the global continuum flux density spectrum of M33 over a very large wavelength range from radio to UV wavelengths. In the course of this analysis, we have demonstrated the importance of accurately accounting for the contribution of CMB fluctuations to the flux density spectrum, which if neglected, results in an over-estimate of $T_\mathrm{dust}$ of $\sim$5\,K and and under-estimate of $\beta_\mathrm{eff}$ of~$\sim$0.4.

Surprisingly, we find that the global integrated emission of M33 between~$\sim$100\,GHz and 3\,THz is adequately described by a single modified blackbody curve, with a mean dust temperature $T_{dust}$ = 21.67~$\pm$~0.30\,K and a mean effective dust emissivity $\beta_{eff}$ = 1.35~$\pm$~0.10, even though such constancy of emission over all of the galaxy and throughout the line of sight is physically unlikely. In order to investigate this, we split M33 into the three independent annuli that the available resolution allows. We find that \textit{both} $T_\mathrm{dust}$ and $\beta_\mathrm{eff}$ decrease from the centre to the outskirts of M33. This correlation is not due to any observational effect and it confirms in a direct manner an earlier conclusion reached by~\citet{Tabatabaei:14}. The dust emission spectrum between~$\sim$100\,GHz and 3\,THz of each of the three sub-regions can be fitted with a single (but not identical) modified blackbody curve, and as the sum of the three curves (the global emission curve) itself is well-fitted with a single modified blackbody curve, we conclude that the combination of individual flux density spectra representing different parts of M33 is highly degenerate with the mean flux density spectrum.

Comparing the global far-IR emission of M33 to that of the other Local Group galaxies for which coverage is available, we find that M33 resembles the Magellanic Clouds rather than the larger spirals, the Milky Way and M31. Within this limited sample, there is a good correlation between the observed Rayleigh-Jeans slope, $\beta_\mathrm{eff}$, and the metallicity of these galaxies, with $\beta_\mathrm{eff}$ increasing with increasing metallicity. This global correlation for the Local Group galaxies is further strengthened by the finding that it also holds \textit{within} M33. The internal M33 $\beta_\mathrm{eff}$ gradient follows its metallicity gradient, with the inner part of M33 much like the LMC, and the outer part much like the lower-metallicity SMC.

%%%%%%% Acknowledgments %%%%%%%%%%%%%%%%%%%%%%%%%%%%%%%%%%%%%%%%%%%%

\section*{Acknowledgments}
We thank the anonymous referee for providing detailed comments that have improved the content of this paper.
MWP acknowledges grant \#2015/19936-1, S\~{a}o Paulo Research Foundation (FAPESP).

%%%%%%%%%%%%%%%  BIBLIOGRAPHY  %%%%%%%%%%%%%%%

%%%%%%%%%%%%%%%%%%%%%%%%%%%%%%%%%%%%%%%%%%%%%%%%%%%%%%%%%%%%%%%%%%%%

\bsp % ``This paper has been produced using the ...''

\label{lastpage}

\end{document}